\documentclass[ba]{imsart}
\pubyear{0000}
\volume{00}
\issue{0}
\doi{0000}
\arxiv{2010.00000}
\firstpage{1}
\lastpage{1}

\usepackage{amsthm}
\usepackage{amsmath}
\usepackage{natbib}
\usepackage[colorlinks,citecolor=blue,urlcolor=blue,filecolor=blue,backref=page]{hyperref}
\usepackage{graphicx}

\startlocaldefs
\numberwithin{equation}{section}
\theoremstyle{plain}

\endlocaldefs

\def\C {\,|\,}

\def \s2{\sigma^2}

\newcommand{\bi}{\begin{itemize}}
\newcommand{\ib}{\end{itemize}}

\newcommand{\be}{\begin{enumerate}[(i)]}
\newcommand{\eb}{\end{enumerate}}

\begin{document}

\begin{frontmatter}
\title{mBART: Multidimensional Monotone BART}

\runtitle{mBART: Multidimensional Monotone BART}

\begin{aug}

\author{\fnms{Hugh A.} \snm{Chipman}\thanksref{addr1}\ead[label=e1]{hugh.chipman@acadiau.ca}},
\author{\fnms{Edward I.} \snm{George}\thanksref{addr2}\ead[label=e2]{edgeorge@upenn.edu}},
\author{\fnms{Robert E.} \snm{McCulloch}\thanksref{addr3}\ead[label=e3]{robert.mcculloch@asu.edu}}
\\ \and
\author{\fnms{Thomas S.} \snm{Shively.}\thanksref{addr4}\ead[label=e4]{Tom.Shively@mccombs.utexas.edu}}

\runauthor{Chipman et al.}

\address[addr1]{Department of Mathematics and Statistics, Acadia University, Nova Scotia, Canada
    \printead{e1} 
}

\address[addr2]{Department of Statistics, The Wharton School, University of Pennsylvania, Philadelphia, PA, U.S.A.
    \printead{e2}
}

\address[addr3]{The School of Mathematical and Statistical Sciences, Arizona State University, Tempe, AZ, U.S.A.
    \printead{e3}
}

\address[addr4]{Department of Information, Risk, and Operations Management, University of Texas at Austin,
Austin, TX, U.S.A.
    \printead{e4}
}


\end{aug}

\begin{abstract}
For the discovery of regression relationships between $Y$ and a large set of $p$ potential predictors $x_1,\ldots,x_p$, the flexible nonparametric nature of BART (Bayesian Additive Regression Trees) allows for a much richer set of possibilities than restrictive parametric approaches. However, subject matter considerations sometimes warrant a minimal assumption of monotonicity in at least some of the predictors. For such contexts, we {  introduce mBART, a constrained version of BART that can flexibly  incorporate monotonicity in any predesignated subset of predictors} using a multivariate basis of monotone trees, while avoiding the further confines of a full parametric form.  { For such monotone relationships, mBART provides} 
(i) function estimates that are smoother and more interpretable,
(ii) better out-of-sample predictive performance,
and (iii) less {post-data} uncertainty.
While many key aspects of the unconstrained BART
model carry over directly to mBART,
the introduction of  monotonicity constraints 
necessitates a fundamental rethinking of how the model is implemented.
In particular, {the original BART Markov Chain Monte Carlo algorithm} relied on a conditional
conjugacy that is no longer available in a {monotonically} constrained space. Various simulated and real examples demonstrate the wide ranging potential of mBART.
\end{abstract}



\end{frontmatter}

\section{Introduction}\label{sec:intro}

Suppose one would like to learn how $Y$ depends on a vector of potential predictors $x = (x_1,\ldots,x_p)$ when no information is available about the form of the relationship. In the absence of such prior information, the Bayesian nonparametric approach BART (Bayesian Additive Regression Trees) can quickly discover the nature of this relationship; see Chipman, George, and McCulloch (2010), hereafter CGM10.  More precisely, based only on the assumption that
\begin{equation}
Y=f(x) + \epsilon, \qquad \epsilon \sim N(0,\sigma^2),
\label{basemodel}
\end{equation}
BART can quickly obtain full posterior inference for the unknown regression function,
\begin{equation}\label{fdef}
f(x) = E(Y \C x)
\end{equation}
and the unknown variance $\sigma^2$.  BART also provides predictive inference as well as model-free variable selection and interaction detection, see Chipman, George, and McCulloch (2013),  Bleich et al. (2014), and Kapelner and Bleich (2016).  Frequentist theoretical support for the attractive empirical performance of BART has been recently developed in Rockova and van der Pas (2020) and Rockova and Saha (2019), and for a kernel-smoothed variant of BART in Linero and Yang (2018).  For an excellent overview of BART and many of its recent related developments, see Hill, Linero and Murray (2020) and the references therein.

{  While the assumption free nature of BART is particularly valuable when a trustable parametric form is unavailable, subject matter considerations sometimes warrant a minimal prior assumption of monotonicity in at least some of the predictors in $x$.  For example, in one of our subsequent illustrative data sets,  $Y$ is the price of a used car and the  $x$ predictors include its age and mileage.  All other things being equal, a prior assumption here that older cars as well as higher mileage cars sell for less on average, is compelling.   Many other contexts where such prior monotonicity assumptions arise naturally, such as dose-response function estimation in epidemiology or market demand function estimation in economics, can be found in the references below.  To harness such monotonicity information, the main goal of this paper is the introduction of monotone BART (hereafter mBART), a constrained version of BART that restricts attention to regression functions $f$ that are monotone in any predesignated subset of the components of $x$, while leaving the remaining components unconstrained}.

In the now rich literature on monotone function estimation, also known as isotonic regression, a wide variety of approaches have been proposed and applied both from the frequentist and Bayesian points of view.  Including constrained nonparametric maximum likelihood, spline modeling, Gaussian processes  {  and projection-based methods} among others, see for example, Barlow et.~al.~(1972), Mammen (1991), Lavine and Mockus (1995),  Ramsay (1998), Holmes and Heard (2003), Neelon and Dunson (2004),  Kong and Eubank (2006), Cai and Dunson (2007), { Chernozhukov, Fernandez-Val and Galichon~(2009)}, Shively, Sager and Walker (2009),  Meyer, Hackstadt and Hoeting (2011), Shively, Walker and Damien (2011), Saarela and Arjas (2011),  Lin and Dunson (2014), Chen and Samworth (2016), Wang and Berger (2016), Lenk and Choi (2017), Wang and Welch (2018), { Lin, St.Thomas, Piegorsch, Scott and Carvalho~(2019), Westling, van der Laan and Carone~(2020)} and the many references therein.   {   In contrast to all these approaches, mBART is built on an easily constrained sum-of-trees approximation of $f$, composed of simple multivariate basis elements that can adaptively incorporate numerous predictors as well as their interactions.  Inheriting the attractive properties of BART, mBART can quickly detect low dimensional signals in high dimensional regression settings with a rapidly mixing MCMC implementation that }generates fully Bayesian uncertainty quantification as its output. 

The extension of BART to our monotonically constrained setting essentially requires two basic innovations.  First, it is necessary to develop general constraints for regression tree functions to be monotone in any predesignated set of coordinates.  Under these constraints, the monotonicity of the full sum-of-trees approximation follows directly.  The second innovation requires a new approach for MCMC posterior computation.  Whereas the original BART formulation allowed straightforward marginalization over regression tree parameters exploiting conditionally conjugate priors, the constrained trees formulation requires a more nuanced approach because complete conjugacy is no longer available.

The outline of the paper is as follows.  
In Section 2, we describe in detail the constrained sum-of-trees model used for monotone function estimation. 
Section 3 discusses the regularization prior for the trees while section 4 describes the new 
MCMC algorithm required to implement mBART. 
Section 5 provides three simulated and two real data examples which illustrate the potential inferential improvements that mBART offers.  Section 6 contains some concluding discussion.

\section{A Monotone Sum-of-Trees Model}\label{sec:constraints}

\begin{figure}
\centerline{\includegraphics[scale= .25]{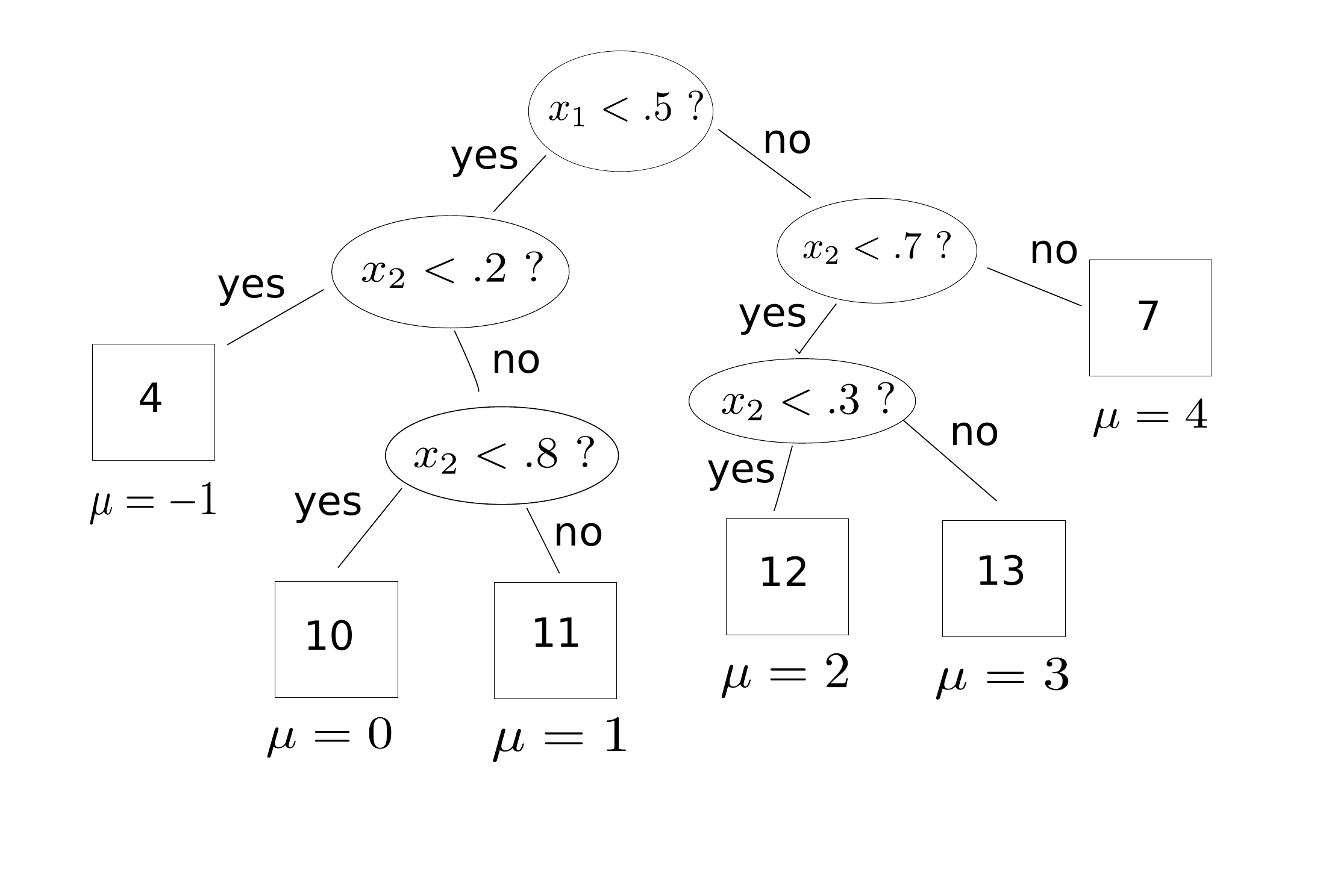}}
\vspace*{-.5in}
\caption{\small A bivariate, monotone regression tree $T$ with 6 terminal nodes.   
Intermediate nodes are labeled with their splitting rules.  
Terminal nodes (bottom leaf nodes) are labeled with their node number.
Below each terminal node is the value of $\mu \in M$ assigned to $x$ by $g(x; T, M)$.}
\label{fig1}
\end{figure}

The essence of BART is a sum-of-trees model approximation of the relationship between $y$ and $x$ in (\ref{basemodel}); 
\begin{equation}\label{sstmodel}
Y = \sum_{j=1}^m g(x; T_j,M_j)\,  + \,
\epsilon,
 \qquad \epsilon \sim N(0,\sigma^2),
\end{equation}
where each $T_j$ is a binary regression tree with a set $M_j$ of associated terminal node constants $\mu_{ij}$, and $g(x; T_j,M_j)$ is the function which assigns $\mu_{ij} \in M_j$ to $x$ according to the sequence of decision rules in $T_j$.    These decision rules are binary partitions of the predictor space of the
form $\{x \le a\}$ vs $\{x > a\}$ where the splitting value $a$ is in
the range of $x$.  (A clarifying example of how $g$ works appears in Figure \ref{fig1} below and is described later in this section).  When $m = 1$, (\ref{sstmodel}) reduces to the single tree model used by 
Chipman et al. (1998)
 for Bayesian CART.
 
Under (\ref{sstmodel}), $E(Y \C x)$  is the sum, over trees $T_1,\ldots, T_m$, of all the terminal node
$\mu_{ij}$'s assigned to $x$ by the $g(x; T_j,M_j)$'s.  
As the $\mu_{ij}$ can take any values
it is easy to see that 
the sum-of-trees model (\ref{sstmodel}) is a flexible representation capable of representing a wide class of functions from $R^n$ to $R$, especially when the number of trees $m$ is large.  Composed of simple functions from $R^p$ to $R$, namely the $g(x; T_j,M_j)$, {  the multivariate step function nature of each tree component greatly facilitates the simple additive imposition of monotone constraints in multiple selected dimensions as described below.  In this way, the sum-of-trees representation is much more manageable than a {multivariate} monotone representation with more complicated basis elements such as multidimensional wavelets or splines, which are often successfully used to more efficiently estimate smooth regression surfaces in low dimensions.  Lastly, because each tree function $g$ is invariant to monotone transformations of $x$ {(with their splitting values)}, predictor standardization choices are not needed for mBART applications. }
 
Key to the construction of mBART are the conditions under which the underlying sum-of-trees function $\sum_{j=1}^m g(x; T_j,M_j)$ will satisfy the following precise definition of a multivariate monotone function. 

{\em Definition:}  For a subset $S$ of the coordinates of $x \in R^n$, a function $f: R^n \rightarrow R$ is said to be {\it monotone in} $S$ if for each $x_i \in S$ and all values of $x$,  $f$ satisfies
\begin{equation}
f(x_1, \ldots, x_i + \delta, \ldots, x_p) \ge f(x_1, \ldots, x_i, \ldots, x_p),
\label{up}
\end{equation}
for all $\delta >0$ ($f$ is nondecreasing), or for all $\delta < 0$ ($f$ is nonincreasing). 

Clearly, a sum-of-trees function will be monotone in $S$ whenever each of the component trees is monotone in $S$.  
Thus it suffices to focus on the conditions for a single tree function $g(x; T, M)$ to be monotone in $S$.  
As we'll see, this will only entail providing constraints on the set of terminal node constants $M$; constraints determined by the tree $T$.  

We illustrate these concepts with the bivariate monotone tree function in Figure 1. 
This tree has six terminal nodes, labeled 4,10,11,12,13, and 7.
The labels follow the standard tree node labeling scheme where the top node is labeled 1
and any non-terminal node with label $j$ has a left child with label $2j$ and a right child with label $2j+1$.
Beginning at the top node, each $x = (x_1,x_2)$  is assigned to subsequent nodes 
according to the sequence of splitting rules it meets.  
This continues until $x$ reaches a terminal node where $g(x; T, M)$ assigns the designated value of $\mu$ from the set $M$.  
For example, with this choice of $(T,M)$, $g(x; T, M)$ = 3 when $x = (.6,.4)$.

Alternative views of the function in Figure 1 are depicted in Figure 2.  
On the left, Figure 2 shows the partitions of the $x$ space induced by $T$.  
The terminal node regions,
$R_4$,$R_{10}$,$R_{11}$,$R_{12}$,$R_{13}$,$R_7$, 
correspond to the six similarly labeled terminal nodes of $T$.
On the right, Figure 2 shows $g(x; T, M)$ as a simple step function which assigns a level $\mu \in M$ to each terminal node region.  From this view, it is clear that for any $x = (x_1,x_2)$, moving $x$ to $(x_1 + \delta, x_2)$ or to $(x_1, x_2 + \delta)$ 
{ cannot decrease} $g$ for $\delta >0$. 
Thus, in the sense of our definition, this $g(x; T, M)$ is monotone in both $x_1$ and $x_2$.

\begin{figure}
\centerline{\includegraphics[scale= .25]{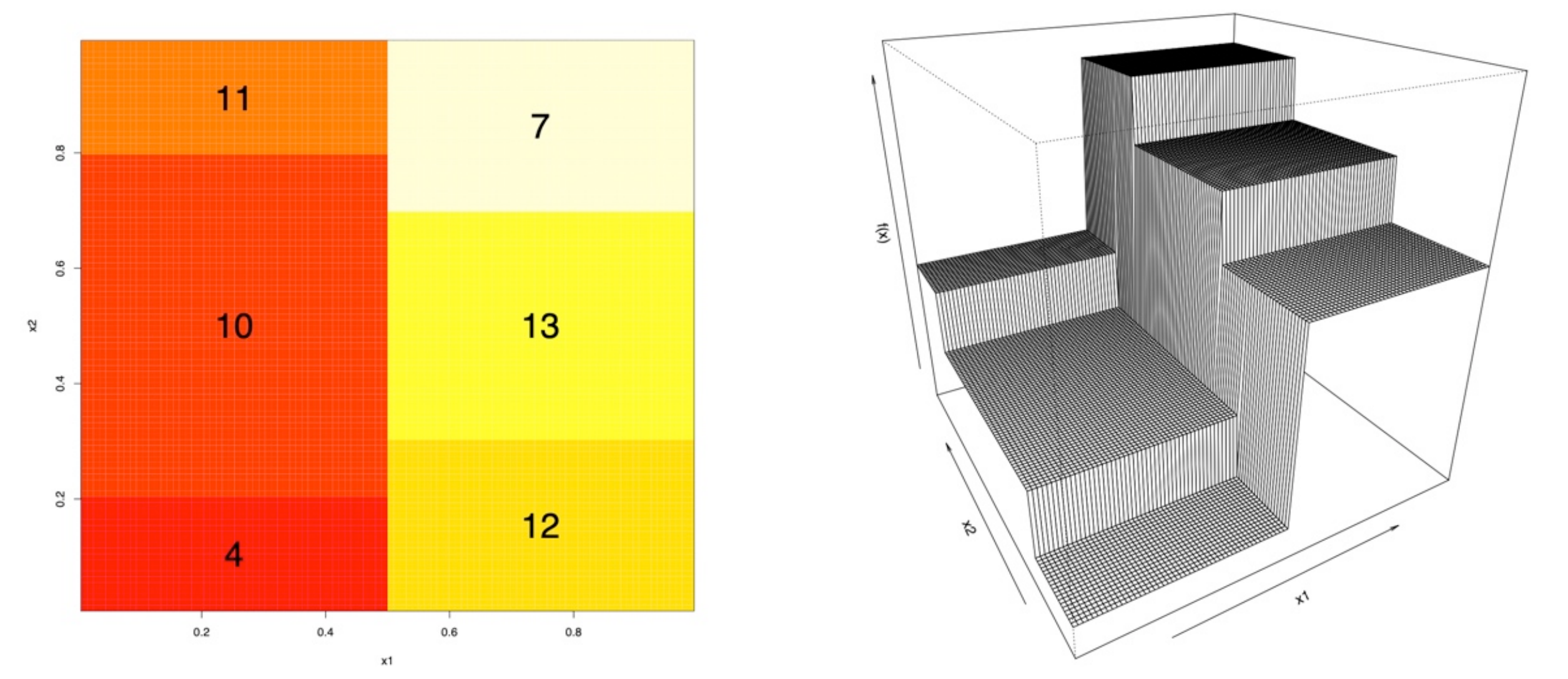}}
\caption{\small
Two alternative views of the bivariate single tree model in Figure 1. 
On the left, the six regions 
$R_4$,$R_{10}$,$R_{11}$,$R_{12}$,$R_{13}$,$R_7$, 
corresponding to the terminal nodes 4,10,11,12,13,7. 
On the right, the levels of the regions assigned by the step function $g(x; T, M)$. }
\label{fig2}
\end{figure}

To see the essence of what is needed to guarantee the monotonicity of a tree function, 
consider the very simple case of a  monotone $g(x; T, M)$ when $T$ is a function of $x=x_1$ only, as depicted in Figure 3.  
Each level region of $g$ corresponds to a terminal node region in $x_1$ space, which is simply an
interval whenever $g$ is a univariate function.  
For each such region, consider the adjoining region with larger values of $x_1$, 
which we refer to as an above-neighbor region, and the adjoining region with smaller values of $x_1$, 
which we refer to as a below-neighbor region.  
End regions will only have single neighboring regions.  
To guarantee (nondecreasing) monotonicity, 
it suffices to constrain the $\mu$ level assigned to each terminal node region 
to be { not greater than} the $\mu$ level of its above-neighbor region, 
and { not less than} the $\mu$  level of its below-neighbor region.

\begin{figure}
\centerline{\includegraphics[scale= .18]{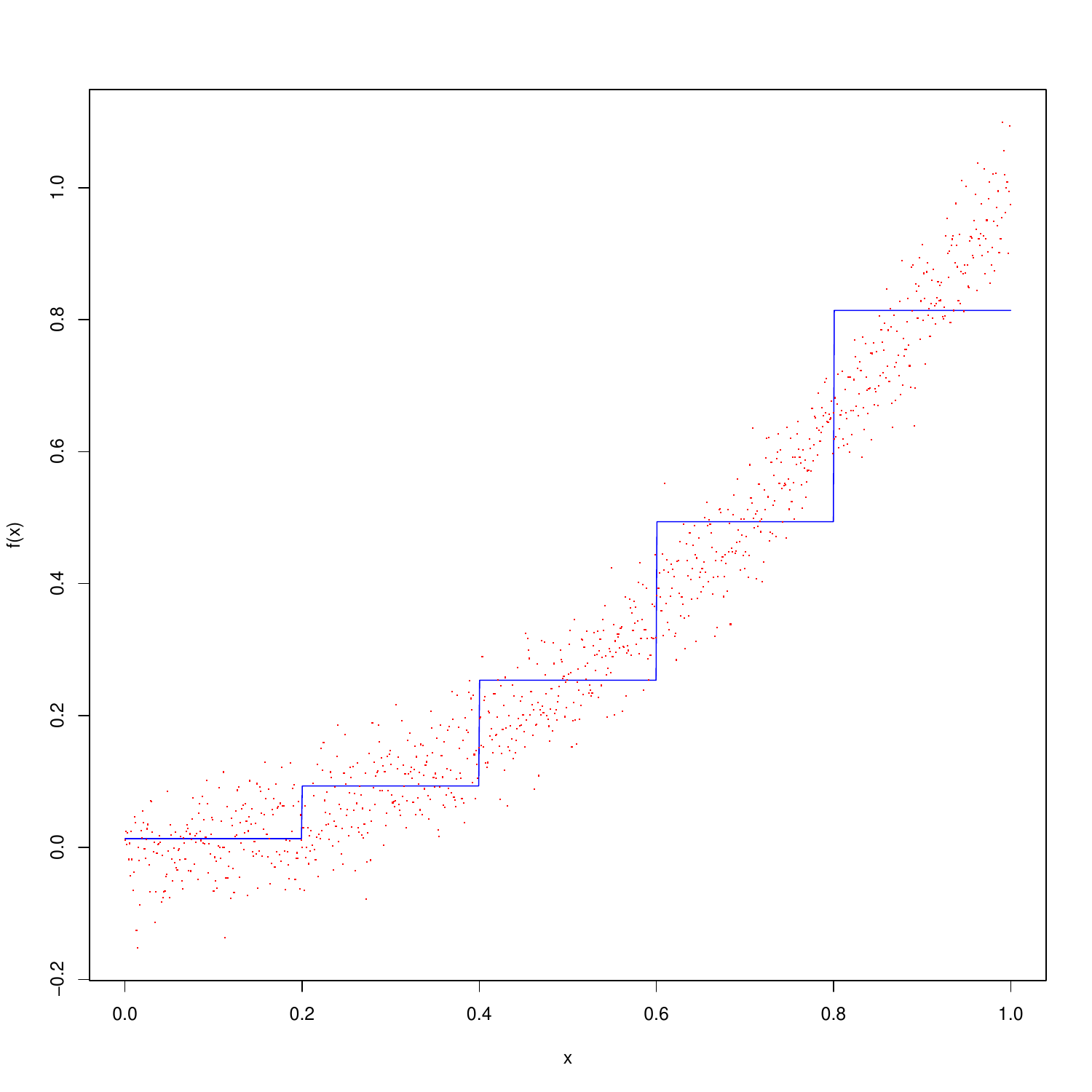}}
\vspace*{-.2in}
\caption{\small A monotone, univariate tree function $g(x; T, M)$.}
\label{fig3}
\end{figure}

To apply these notions to a bivariate tree function $g(x; T, M)$ 
as depicted in Figures 1 and 2, we will simply say that rectangular regions are 
neighboring if they have boundaries which are adjoining in any of the coordinates.  
Furthermore, a region $R_k$ will be called an above-neighbor of a region $R_{k^*}$ 
if the lower  adjoining boundary of $R_k$ is the upper adjoining boundary of $R_{k^*}$.  
A below-neighbor is defined similarly.  
For example, in Figure 2, $R_7$ is an 
above-neighbor of $R_{10},R_{11}$ and $R_{13}$; and $R_{10}$ and $R_{12}$ are below-neighbors of $R_{13}$.  

Note that $R_4$ and $R_{13}$ are not neighbors.
We will say the $R_4$ and $R_{13}$ are {\it separated} because the $x_2$ upper boundary of $R_4$ is less
than the $x_2$ lower boundary of $R_{13}$.  
For a small enough step size $\delta$, it is impossible to get from
$R_4$ to $R_{13}$
by changing any $x_i$ by $\delta$
so that the mean level of one does not constrain the mean level of the other.

To make these definitions precise for a $d$-dimensional tree $T$ (a function of $x = (x_1,\ldots, x_d)$), 
we note that each terminal node region of $T$ will be a rectangular region of the form
\begin{equation}
R_k = \{x :  x_i \in [L_{ik},U_{ik}), i = 1,\dots, d\},
\end{equation}
where the interval $[L_{ik},U_{ik})$ for each $x_i$ is determined by the sequence of splitting rules leading to $R_k$.  

We say that $R_k$ is separated from $R_{k^*}$ if $U_{ik} < L_{ik^*}$
or $L_{ik} > U_{ik^*}$ for some $i$.
In Figure 2, $R_{13}$ is separated from $R_4$ and $R_{11}$.

If $R_k$ and $R_{k^*}$ are not separated,
$R_k$ will be said to be an above-neighbor of $R_{k^*}$ if $L_{ik} = U_{ik^*}$ for some $i$, 
and it will be said to be a below-neighbor of $R_{k^*}$ if $U_{ik} = L_{ik^*}$ for some $i$.  
Note that any terminal node region may have several above-neighbor and below-neighbor regions.  
$R_{13}$ has below neighbors $R_{10}$ and $R_{12}$ and above neighbor $R_7$.

The constraints on the $\mu$ levels under which $g(x; T, M)$ will be monotone are now straightforward to state.  

{ {\it Constraint Conditions for Tree Monotonicity}:  A tree function $g(x; T, M)$ will be monotone in coordinate $x_i$ if the $\mu$ level of each of its terminal node regions is\\ 
(a) not greater than the minimum level of all of its above-neighbor regions in the $x_i$ direction, and \\
(b) not less than the maximum level of all of its below-neighbor regions in the $x_i$ direction.}  

The function $g$ will be monotone in  $S$ if the neighboring regions satisfy (a) and (b) for all the coordinates in $S$
(rather than all coordinates).

As we'll see in subsequent sections, an attractive feature of these conditions is that they dovetail perfectly with the nature of our iterative MCMC simulation calculations.  At each step there, we simulate one terminal node level at time conditionally on all the other node levels, so imposing the constraints is straightforward.  This avoids the need to simultaneously constrain all the levels across all trees at once.  

\section{A Constrained Regularization Prior}\label{sec:regprior}

The mBART model specification is completed by putting a constrained regularization prior on the parameters, $(T_1,M_1),\ldots,(T_m,M_m)$ and $\sigma$, of the sum-of-trees model \eqref{sstmodel}.   Essentially a modification of the original BART prior formulation to accommodate {monotone constraints in a predesignated subset $S$ of the coordinates of $x$}, we follow CGM10 and proceed by restricting attention to priors of the form
\begin{eqnarray}\label{eq:mainform}
p((T_1,M_1),\ldots,(T_m,M_m),\sigma)
= \left[\prod_j p(M_j \C T_j)\,p(T_j)\right] \,
 p(\sigma)\label{indep1},
\end{eqnarray}
where the tree components $(T_1,M_1),\ldots,(T_m,M_m)$ are apriori independent of each other and of $\sigma$.  

As discussed in the previous section, a sum-of-trees function  $\sum_{j=1}^m g(x; T_j,M_j)$ is guaranteed to be monotone in $S$  whenever each of the trees $g(x; T_j,M_j)$ is monotone for each $x_i$ in $S$ in the sense of \eqref{up}.   Thus, it suffices to restrict the support of $p(M_j \C T_j)$ to $\mu_{ij}$ values which satisfy the Monotonicity Constraints (a) and (b) from Section \ref{sec:constraints}.   For this purpose, let $C$ be the set of all $(T,M)$ which satisfy these monotonicity constraints, namely
\begin{equation}\label{Cdef}
C = \{(T,M): g(x; T, M) \mbox{ is monotone in $x_i \in S$} \}.
\end{equation}
These constraints are then incorporated into the prior by constraining the CGM10 BART independence form
$p(M_j \, | \, T_j) = \prod_i{p(\mu_{ij} \C T_j)}$ to have support only over $C$,
\begin{equation}\label{eq:general}
p(M_j \, | \, T_j) \propto \left[\prod_{i=1}^{b_j} p(\mu_{ij} \C T_j)\right] \, \chi_C (T_j,M_j).
\end{equation}
Here $b_j$ is the number of bottom (terminal) nodes of $T_j$, 
and $\chi_C (\cdot) = 1$ on $C$ and $=0$ otherwise.  {  The effect of this prior is to directly constrain the support of the posterior distribution to those sum-of-tree functions comprised only of components in $C$.}   



In the next three subsections we discuss the choice of priors
$p(T_j)$, $p(\sigma)$, and $p(\mu_{ij} \C T_j)$.
These will have the same form as in CGM10, but in
some cases the monotonicity constraint will motivate modifications
for our recommended hyperparameter settings.

\subsection{Calibrating the $T_j$ Prior}\label{sec:treeprior}

The tree prior $p(T_j)$ is specified by three aspects: (i) the probability of a node having children at
depth $d$ ($= 0, 1,2,\ldots$) is 
\begin{equation}
\alpha (1+d)^{-\beta}, \qquad
\alpha \in (0,1), \beta \in [0, \infty),
\label{treeprior}
\end{equation}
(ii) the uniform distribution over available predictors for splitting rule assignment at
each interior node, and (iii) the uniform distribution on the discrete set of available splitting
values for the assigned predictor at each interior node.   This last choice has the appeal of invariance
under monotone transformations of the predictors.  

Because we want the
regularization prior to keep the individual tree components small, 
especially when $m$ is set to be large,
we typically recommend the defaults
$\alpha=.95$ and $\beta=2$ in (\ref{treeprior}) in the unconstrained case. 
With this choice, simulation of tree skeletons directly from (i) shows us that 
trees with 1, 2, 3, 4, and $\geq 5$ terminal nodes will receive prior
probabilities of about 0.05, 0.55, 0.28, 0.09, and 0.03, respectively.

Discussion of the choice of $\alpha$ and $\beta$ in the constrained case
is deferred to the end of Section~\ref{sec:impMH}
since our choices are motivated by details of the Markov Chain Monte Carlo
algorithm for posterior computation.

\subsection{Calibrating the $\sigma$ Prior}\label{sec:sigmaprior}

For $p(\sigma)$, we use the (conditionally) conjugate inverse chi-square distribution
$\sigma^2 \sim \nu \, \lambda/\chi_{\nu}^2$.  To guide the specification
of the hyperparameters $\nu$ and $\lambda$, we recommend a data-informed approach to assign substantial probability to the entire region of
plausible $\sigma$ values while avoiding overconcentration and overdispersion.
This entails calibrating the prior degrees of freedom $\nu$ and scale $\lambda$ 
using a ``rough data-based overestimate''
$\hat{\sigma}$ of $\sigma$.  

The two natural choices for $\hat{\sigma}$ are (1) the ``naive''
specification, in which we take $\hat{\sigma}$ to be the sample
standard deviation of $Y$ (or some fraction of it), or (2) the ``linear model''
specification, in which we take $\hat{\sigma}$ as the residual
standard deviation from a least squares linear regression of $Y$
on the original $x$'s.  We then pick a value of
$\nu$ between 3 and 10 to get an appropriate shape, and a value of
$\lambda$ so that the $q$th quantile of the prior on $\sigma$ is located at
$\hat{\sigma}$, that is $P(\sigma < \hat{\sigma}) = q.$  We consider values of $q$ such as
0.75, 0.90 or 0.99 to center the distribution below $\hat{\sigma}$.
For automatic use, we recommend the
default setting $(\nu, q) = (3, 0.90)$ which tends to avoid
extremes.  Alternatively, the values of $(\nu, q)$ may be chosen by
cross-validation from a range of
reasonable choices.
This choice is exactly as in CGM10.

{  An advantage of this data-informed approach to the calibration of $p(\sigma)$ is that it allows for semi-automatic ``off-the-shelf'' implementations with selected tuning parameters.  However, an expert with reliable prior information could use this same scheme but with $\hat{\sigma}$ obtained as a subjective estimate  of a selected $q$th quantile of $\sigma$, thereby avoiding any need to use the data for this purpose.}

\subsection{Calibrating the $M_j\C T_j$ Prior}\label{sec:mprior}

{  For the choice of $p(\mu_{ij} \C T_j)$ in \eqref{eq:general}, we adopt normal densities as used in BART, but now with different prior variance choices depending on whether {or not} $\mu_{ij}$ is constrained by the set $C$ in \eqref{Cdef}.  For $\mu_{ij}$ unconstrained by $C$, we use a $N(\mu_\mu, \sigma_\mu^2)$ prior so that
\begin{equation} \label{eq:unconstrained} 
p(\mu_{ij} \, | \, T_j) = \phi_{\mu_\mu, \sigma_\mu},
\end{equation}
the normal density with mean $\mu_\mu$ and variance $\sigma_\mu^2$.
However for $\mu_{ij}$ constrained by $C$, we use a $N(\mu_\mu, c^2\sigma_\mu^2)$ prior with the choice $c^2= \frac{\pi}{\pi-1} \approx 1.4669$ so that 
\begin{equation} \label{eq:constrained} 
p(\mu_{ij} \, | \, T_j) = \phi_{\mu_\mu, c\sigma_\mu}. 
\end{equation}

To motivate the increased variance choice in \eqref{eq:constrained}, consider a simple tree with just two terminal node means $\mu_1$ and $\mu_2$ constrained to satisfy $\mu_1 \le \mu_2$.  Under \eqref{eq:constrained} with this constraint, the joint distribution of $\mu_1$ and $\mu_2$ is
\begin{equation}\label{eq:2nodedist}
p(\mu_1,\mu_2) \propto \phi_{\mu_\mu, c\sigma_\mu}(\mu_1) \phi_{\mu_\mu, c\sigma_\mu}(\mu_2) \, \chi_{\{\mu_1 \le \mu_2\}}(\mu_1,\mu_2).
\end{equation}
Integrating each of $\mu_1$ and $\mu_2$ out from $p(\mu_1,\mu_2)$, yields the marginal distributions of  $\mu_1$ and $\mu_2$,
\begin{eqnarray}
p(\mu_1) & \propto & \phi_{\mu_\mu, c\sigma_\mu}(\mu_1) \Phi_{\mu_\mu, c\sigma_\mu}(-\mu_1) \\
p(\mu_2) & \propto & \phi_{\mu_\mu, c\sigma_\mu}(\mu_1) \Phi_{\mu_\mu, c\sigma_\mu}(\mu_2).
\end{eqnarray}
These are skew normal distributions which, when $c^2 = \frac{\pi}{\pi-1}$, have the same variances $\sigma_\mu^2$ and respective means $\mu_\mu - \sigma_\mu/\sqrt{\pi-1}$ and  $\mu_\mu + \sigma_\mu/\sqrt{\pi-1}$, (Azzalini 1985).  That the prior variances of the constrained means $\mu_1$ and $\mu_2$ match the prior variances of the unconstrained means in \eqref{eq:unconstrained},  helps to balance the prior effects across predictors and facilitates the calibrated specification of $\sigma_\mu$ described below.  Of course, it will be occasionally the case that some means $\mu_{ij}$ may be further constrained when they occur deeper down the tree, thereby further reducing their prior variance.  Although additional small prior adjustments can be considered for such cases, we view them as relatively unimportant because the vast majority of BART trees will be small with at most one or two constraints.  Thus, we adopt the prior \eqref{eq:constrained} for any $\mu_{ij}$ which becomes constrained}.

To guide the specification of the hyperparameters $\mu_\mu$ and $ \sigma_\mu$, we use the same informal empirical Bayes strategy in CGM10. Based on the idea that that $E(Y \C x)$ is very likely between $y_{min}$ and $y_{max}$, the observed minimum and maximum of $Y$, we want to choose $\mu_\mu$ and $\sigma_\mu$ so that the induced prior on  $E(Y \C x)$ assigns substantial probability to the interval
$(y_{min}, y_{max})$.  By using the observed $y_{min}$ and $y_{max}$, we aim to ensure that the implicit prior for $E(Y \C x)$ is in the right ``ballpark''{ , thereby avoiding prior-data conflict.}  

In the unconstrained case where each value of $E(Y \C x)$ is the sum of $m$ iid $\mu_{ij}$'s under the sum-of-trees model, 
the induced prior on $E(Y \C x)$ under \eqref{eq:unconstrained}  is exactly $N(m\, \mu_\mu, m\, \sigma_\mu^2)$.  
Let us argue now that when monotone constraints are introduced, $N(m\, \mu_\mu, m\, \sigma_\mu^2)$ 
still holds up as a useful approximation to the induced prior on $E(Y \C x)$.   
To begin with, for each value of $x$, let $g(x; T_j,M_j) = \mu_{xj}$, the mean assigned to $x$ by the $j$th tree $T_j$. 
Then, under the sum-of-trees model, $E(Y \C x) = \sum_{j=1}^m \mu_{xj}$ is the 
sum of $m$ independent means since the $\mu_{xj}$'s are independent across trees.  
Using central limit theorem considerations, this sum of small random effects will be approximately 
normal, at least for the central part of the distribution.  
The means of all the random effects will be centered around $\mu_\mu$, 
(the constrained $\mu_{ij}$'s will have pairwise offsetting biases), 
and so the mean of $E(Y \C x)$ will be approximately $\mu_\mu$.  
Finally, since the marginal variance for all $\mu_{xj}$'s \ is at least approximately $\sigma_\mu^2$, 
the variance of $E(Y \C x)$ will be approximately $m\sigma_\mu^2$.  

Proceeding as in CGM10, we thus choose $\mu_\mu$ and $ \sigma_\mu$ so that $m\, \mu_\mu -  k\, \sqrt{m}\, \sigma_\mu = y_{min}$ and  $m \, \mu_\mu +  k\, \sqrt{m}\, \sigma_\mu = y_{max}$ for some preselected value of $k$. This is conveniently implemented by first shifting and rescaling $Y$ so that the observed transformed $y$ values range from  $y_{min}= -0.5$ to $y_{max}= 0.5$, and then setting $\mu_\mu = 0$ and $\sigma_{\mu} = 0.5/k \sqrt{m}$. 
Using $k=2$, for example, would yield a 95\% prior probability that $E(Y \C x)$ {  over the range of $x$} is in the interval $(y_{min}, y_{max})$, thereby assigning substantial probability to the entire region of plausible values of $E(Y \C x)$ while avoiding overconcentration and overdispersion.   As $k$ and/or the number of trees $m$ is increased, this prior will become tighter, thus limiting the effect of the individual tree components of (\ref{sstmodel}) by keeping the $\mu_{ij}$ values small. We have found that values of $k$ between 1 and 3 yield good results, and we recommend $k = 2$ as an automatic default choice, the same default recommendation for BART.  Alternatively, the value of $k$ may be chosen by cross-validation from a range of reasonable choices. 

{  Just as for the calibration of $p(\sigma)$ above, an advantage of this data-informed approach to the calibration of $p(\mu_{ij} \C T_j)$ is that it allows for semi-automatic ``off-the-shelf'' implementations with selected tuning parameters.  Here too, however, an expert with reliable prior information could use this same scheme with a subjective estimate of an interval which will contain $E(Y \C x)$ over the range of $x$ with high probability, thereby completely avoiding the need to use the data for this purpose.  We illustrate how this can be carried out with real expert input in the stock return application in Section \ref{subsec:fin}.
}

{
\subsection{The Choice of $m$}\label{sec:numtrees}
}
Again as in BART, we treat $m$ as a fixed tuning constant to be chosen by the user.  
For prediction, we have found that mBART performs well with values of at least $m = 50$.  
For variable selection, values as small as $m = 10$ are often effective. 

\vspace{.5cm}
\section{MCMC Simulation of the Constrained Posterior} 

\subsection{Bayesian Backfitting of Constrained Regression Trees}

Let $y$ be the $n\times 1$ vector of independent observations of $Y$ from \eqref{sstmodel}. 
All post-data information for Bayesian inference about any aspects of the unknowns, 
$(T_1,M_1),\ldots,(T_m,M_m)$, $\sigma$ and future values of $Y$, 
is captured by the full posterior distribution 
\begin{equation}\label{posterior2}
p((T_1,M_1), \ldots,(T_m,M_m),\sigma \C y). 
\end{equation}
Since all inference is conditional on the predictor $x$ values, we suppress them in the notation.
This posterior is proportional to the product of the likelihood
$p(y \C (T_1,M_1),\ldots,(T_m,M_m), \sigma)$, which is the product of normal likelihoods based on \eqref{sstmodel}, 
and the constrained regularization prior $p((T_1,M_1),\ldots,(T_m,M_m),\sigma)$  described in Section \ref{sec:regprior}. 

To extract information from \eqref{posterior2}, which is generally intractable, we propose an MCMC backfitting algorithm that simulates a sequence of draws, $k = 1,\dots, K$, 
\begin{equation}\label{draws}
(T_1,M_1)^{(k)}, \ldots,(T_m,M_m)^{(k)},\sigma^{(k)} 
\end{equation}
that is converging in distribution to \eqref{posterior2} as $K \rightarrow \infty$.

Beginning with a set of initial values of $((T_1,M_1)^{(0)}, \ldots,(T_m,M_m)^{(0)},\sigma^{(0)})$, 
{the outer loop of this algorithm proceeds as in CGM10 by simulating} a sequence of transitions 
$(T_j,M_j)^{(k)}$ $\rightarrow$ $(T_j,M_j)^{(k+1)}$, 
for $j = 1,\ldots, m$,
$\sigma^{(k)} \rightarrow \sigma^{(k+1)}$. 
The 
$(T_j,M_j)^{(k)}$ $\rightarrow$ $(T_j,M_j)^{(k+1)}$
transition 
is obtained by using a Metropolis-Hastings (MH)  algorithm to simulate a single transition of a Markov chain with stable distribution
\begin{equation} \label{newdraw}
p((T_j,M_j) \C r_j^{(k)}, \sigma^{(k)}),
\end{equation}
for $j = 1,\ldots,m$, where
\begin{equation} \label{Rj}
r_j^{(k)} \equiv y - \sum_{j' < j} g(x;T_{j'},M_{j'})^{(k+1)} - \sum_{j' > j} g(x;T_{j'},M_{j'})^{(k)}
\end{equation}
is the $n-$vector of partial residuals based on a fit that excludes
the most current simulated values of $T_{j'},M_{j'}$ for $j' \ne j$. A full iteration of the algorithm is then completed by simulating the draw of $\sigma^{(k+1)}$ from the full conditional
\begin{equation}
\sigma \C (T_1,M_1)^{(k+1)},\ldots,(T_m,M_m)^{(k+1)}, y \label{drawsig}.
\end{equation}
Because conditioning the distribution of  $(T_j,M_j)$ on $r_j^{(k)}$ and $\sigma^{(k)}$ in \eqref{newdraw} is equivalent to conditioning on the excluded values of
$(T_{j'},M_{j'})$, $\sigma^{(k)}$ and $y$, this algorithm is an instance of MH within a Gibbs sampler.  

\subsection{A New Localized Metropolis-Hastings Algorithm}\label{eq:moves}

{ To accommodate the constrained nature of the prior \eqref{eq:general}, we now introduce} a new localized MH algorithm for the simulation of  
$(T_j,M_j)^{(k)}$ $\rightarrow$ $(T_j,M_j)^{(k+1)}$ as single transitions of a 
Markov chain converging to the (possibly constrained) posterior \eqref{newdraw}.   
For simplicity of notation, let us denote a generic instance of these moves  by $(T^0, M^0)$ $\rightarrow$ $(T^1, M^1)$.  
Dropping $\sigma^{(k)}$ from \eqref{newdraw} since it is fixed throughout this move, 
and dropping all the remaining subscripts and superscripts, 
the target posterior distribution can be expressed as
\begin{equation} \label{newdraw2}
p(T,M \C r) = p(r \C T, M) p(M \C T) p(T) /p(r),
\end{equation}
where its components are as follows.  

First, $p(r \C T, M)$ is the normal likelihood which would correspond to an observation of 
$r = g(x; T,M) + \epsilon$, where $\epsilon \sim N_n(0, \sigma^2 I)$.  
Assuming $M = (\mu_1,\ldots, \mu_b)$, and letting $r_i$ be the vector of components of 
$r$ assigned to $\mu_i$ by $T$, this likelihood is of the form
\begin{equation}\label{eq:indep-r}
p(r \C T, M) = \prod_{i=1}^b p(r_i \C \mu_i) 
\end{equation}
where
\begin{equation}
p(r_i \C \mu_i)  \propto \prod_j \exp(-(r_{ij} - \mu_i)^2/2\sigma^2).  
\end{equation}
The prior of $M \C T$ given by \eqref{eq:general} is of the form
\begin{equation}\label{eq:general2}
p(M \, | \, T) \propto \left[\prod_{i=1}^{b} p(\mu_i \C T)\right] \, \chi_C (T,M),
\end{equation}
where $p(\mu_i \C T) = \phi_{\mu_\mu, \sigma_\mu}(\mu_i)$ from   \eqref{eq:unconstrained}  
if $\mu_i$ is unconstrained by $\chi_C$, and $p(\mu_i \C T) = \phi_{\mu_\mu, c\sigma_\mu}(\mu_i)$ 
from  \eqref{eq:constrained}  if $\mu_i$ is constrained by $\chi_C$.  
The tree prior $p(T)$ described in Section \ref{sec:treeprior} is the same form used for unconstrained BART. 
Finally, the intractable marginal $p(r)$, 
which would in principle be obtained by summing and integrating 
over $T$ and $M$, will fortunately play no role in our algorithm.

In unconstrained CART and BART, CGM98 and CGM10 used the following two step Metropolis-Hastings 
(MH) procedure for the simulation of $(T^0, M^0)$ $\rightarrow$ $(T^1, M^1)$.   
First, a proposal $T^*$ was generated with probability $q(T^0\rightarrow T^*)$.  
Letting $q(T^*\rightarrow T^0)$ be the probability of the reversed step, 
the move  $T^1 = T^*$ was then accepted with probability 
\begin{eqnarray} \nonumber
\alpha & = & \min \left\{\frac{q(T^*\rightarrow T^0)}{q(T^0\rightarrow T^*)}\, \frac{p(T^* \C r)}{p( T^0 \C r)},\; 1 \right\} \\ \label{eq:alphaU}
& = & \min \left\{\frac{q(T^*\rightarrow T^0)}{q(T^0\rightarrow T^*)}\, \frac{p(r \C  T^*)}{p(r \C  T^0)} \frac{p(T^*)}{p(T^0)},\; 1 \right\} .
\end{eqnarray}
If accepted,  
any part of 
$M^1$ with a new ancestry under $M^1$ is simulated 
from independent normals since 
$p(M \C T^1, r)$ just consists of $b$ independent normals
given the independence and conditional conjugacy of our prior
(which is \eqref{eq:general2} without the monotonicity constraint $\chi_C (T,M)$)
and the conditional data independence \eqref{eq:indep-r}. 
Otherwise $(T^1, M^1)$ is set equal to  $(T^0, M^0)$.  

In the contrained case, the basic algorithm is the same
except that with the monotonicity constraint in \eqref{eq:general2},
the $\mu_i$ in $M$ are dependent.
Hence, when we make local moves involving a few of the $\mu_i$
we must be careful to condition on the remaining elements.
In addition, computations must be done numerically since we lose
the conditional conjugacy.  
The moves in mBART only operate on one or two of the $\mu$ values at a time
so that the appropriate conditional integrals can easily be done numerically.

We consider  localized proposals $(T^0, M^0)$ $\rightarrow$ $(T^*, M^*)$ 
under which $M^0$ and $M^*$ differ only by those 
$\mu$'s which have different ancestries under $T^0$ and $T^*$.  
Letting $\mu_{same}$ be the part of $M^0$ with the same ancestry under $T^0$ and $T^*$, 
we restrict attention to proposals for which $M^0 = (\mu_{same}, \mu_{old})$ and 
$M^* = (\mu_{same}, \mu_{new})$, where $\mu_{old}$ is the part of $M^0$ that 
will be replaced by  $\mu_{new}$ in $M^*$.   
It will also be convenient in what follows to let $r_{old}$ be the components of the data $r$  
assigned to $\mu_{old}$ by $T^0$, $r_{new}$ to be the components assigned to $\mu_{new}$ by $T^*$, 
and $r_{same}$ to be the components assigned to the 
identical components of $\mu_{same}$ by both $T^0$ and $T^*$.  

\begin{figure}
\centerline{\includegraphics[scale= .47]{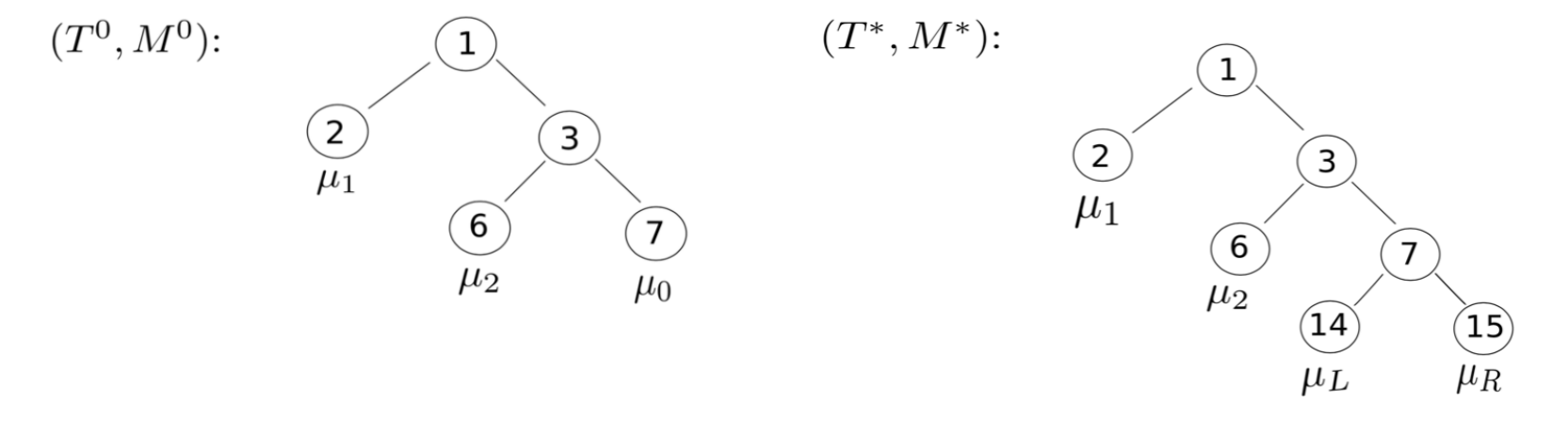}}
\vspace{-.3cm}
\caption{\small
A typical birth step starting at $(T^0, M^0)$ and proposing $(T^*, M^*)$.  
$T^0$ includes the nodes 1,2,3,6,7.
$T^*$ includes the nodes 1,2,3,6,7,14,15.
Here $\mu_{\it same} = (\mu_1,\mu_2)$.
Our MH step proceeds conditionally on $\mu_{\it same}$ 
and the associated ancestral parts of the tree structures $T^0$ and $T^*$, nodes 1,2,3,6.
Our proposal generates the candidate rule associated with node 7 in $T^*$.
Conditional on all these elements, we integrate out $\mu_0$ or $(\mu_L,\mu_R)$ subject
to the constraints implied by the conditioning elements.
Note that the proposal for the node 7 rule does not depend on $\mu_{\it same}$, it only
depends on the tree structures. 
}

\label{figBD}
\end{figure}

For example, suppose we begin with a proposal $T^0\rightarrow T^*$ that randomly 
chooses between a birth step and death step, and that $T^*$ was obtained by a birth step, 
which entails adding two child nodes at a randomly chosen terminal node of $T^0$.   
This move is illustrated in \mbox{Figure \ref{figBD}} where $M^0 = (\mu_1, \mu_2, \mu_{0})$ and 
$M^* = (\mu_1, \mu_2, \mu_L, \mu_R)$, so that $\mu_{same} = (\mu_1, \mu_2)$ to which 
$r_{same} = (r_1, r_2)$ is assigned,  
$\mu_{old} = \mu_0$ to which $r_{old} = r_0$ is assigned, and $\mu_{new} = (\mu_L, \mu_R)$ to which $r_{new} = (r_L, r_R)$ is assigned.  
Note that the set of observations in $(r_L, r_R)$ is just the division of the set of observations in $r_0$
defined by the decision rule associated with node 7 in the tree $T^*$.

The key is to then proceed conditionally on $\mu_{same}$ and the tree ancestry associated with it.
In Figure~\ref{figBD}, we condition on $\mu_{same} = (\mu_1,\mu_2)$ {\it and} the ancestral tree structure
given by nodes $(1,2,3,6)$ 
including the decision rules associated with the interior nodes 1 and 3.
To keep the notation clean, we will use $\mu_{same}$ as a conditioning variable in our expressions
below and the reader must make a mental note to
include the associated tree ancestry as conditioning information.

Conditionally on $\mu_{same}$,
our Metropolis procedure  is as follows.  
First, 
a proposal $T^*$ is generated with probability $q(T^0\rightarrow T^*)$, using the same CGM98 proposal used in unconstrained CART and BART.  
Letting $q(T^*\rightarrow T^0)$ be the probability of the reversed step, the move  $T^1 = T^*$ is then accepted with probability 
\begin{eqnarray} \nonumber
\alpha & = &\min \left\{\frac{q(T^*\rightarrow T^0)}{q(T^0\rightarrow T^*)}\, \frac{p(T^* \C \mu_{same}, r)}{p( T^0 \C \mu_{same}, r)
},\; 1 \right\}\\ \nonumber
&=& \min \left\{\frac{q(T^*\rightarrow T^0)}{q(T^0\rightarrow T^*)}\, \frac{p(T^* \C \mu_{same}, r_{new})}{p( T^0 \C \mu_{same}, r_{old})
},\; 1 \right\} \\ \label{eq:alphaM}
&=& \min \left\{\frac{q(T^*\rightarrow T^0)}{q(T^0\rightarrow T^*)}\,\frac{p(r_{new} \C  T^*, \mu_{same})}{p(r_{old} \C T^0, \mu_{same} )}  \frac{p(T^*)}{p(T^0)},\; 1 \right\}.
\end{eqnarray}
The difference between \eqref{eq:alphaU}  and \eqref{eq:alphaM}
is that we condition on $\mu_{same}$ throughout and explicitly note
that the $r_{same}$ part of $r$ does not matter.
In going from the first line above to the second
we have
used the fact that, 
conditional on  $\mu_{same}$, 
$r_{same}$ gives the same multiplicative contribution to the
top and bottom of the acceptance ratio so that it cancels out
leaving only terms depending on $r_{new}$ and $r_{old}$.
To go from the second line above to the third we will compute the required
$r_{new}$ and $r_{old}$ marginals numerically  as detailed in Section~\ref{sec:impMH} below. 
Note also that in the BART prior, $T$ and $M$ are dependent 
only through the dimension of $M$
so $p(T^*) \; / \; p(T^0)$ is the same as in the unconstrained case.

If $T^1 = T^*$ is accepted, $\mu_{new}$ is then simulated from  
$p(\mu_{new} \C T^1, \mu_{same}, r) = p(\mu_{new} \C T^1, \mu_{same}, r_{new})$ 
and $M^1$ is set equal to $(\mu_{same}, \mu_{new})$.  
Otherwise $(T^1, M^1)$ is set equal to  $(T^0, M^0)$.

\subsection{Implementation of the Localized MH Algorithm}\label{sec:impMH}

The implementation of our localized MH algorithm requires 
the evaluation of $p(r_{new} \C  T^*, \mu_{same})$ 
and  $p(r_{old} \C T^0, \mu_{same})$ for the $\alpha$ calculation in \eqref{eq:alphaM}, 
and the simulation from $p(\mu_{new} \C T^1, \mu_{same}, r_{new})$.   
Although these can all be done quickly and easily in the unconstrained cases, 
a different approach is needed for constrained cases.  
This approach, which we now describe, relies crucially on the reduced computational 
requirements for the localized MH algorithm when $T^0\rightarrow T^*$ 
is restricted to local moves at a single node.  

For the moment, consider the birth move described in 
Section~\ref{eq:moves} and illustrated in \mbox{Figure~\ref{figBD}}. 
In this case, $\mu_{new} = (\mu_L,\mu_R)$ with corresponding  
$r_{new} = (r_L,r_R)$ 
and
$\mu_{old} = \mu_0$ with corresponding $r_0$.
Thus, to perform this move, it is necessary to compute $p(r_L, r_R \C T^*, \mu_{same})$ and 
$p(r_0 \C T^0, \mu_{same})$ for the computation of $\alpha$ in \eqref{eq:alphaM}, 
and to simulate $(\mu_L, \mu_R)$ from
 $p(\mu_L, \mu_R \C r_L, r_R, T^*, \mu_{same})$ when $T^1 = T^*$ is selected.  
For the corresponding death step, we would need to simulate $\mu_0$ from $p(\mu_0 \C r_0, T^0, \mu_{same})$.  
When these means are unconstrained, these calculations can be done quickly with 
closed form expressions and the simulations by routine methods
so we focus here on the constrained case.

Let us begin with the calculation of 
\begin{equation}\label{integral1}
p(r_L, r_R \C T^*, \mu_{same})= \int p(r_L \C \mu_L) \, p(r_R  \C \mu_R) \, 
p(\mu_L, \mu_R \C T^*, \mu_{same}) \, d\mu_L\,d\mu_R
\end{equation} 
where
\begin{equation}\label{eq:dstar}
p(\mu_L, \mu_R \C T^*, \mu_{same}) =  
\phi_{\mu_\mu, c\sigma_\mu}(\mu_L) \, \phi_{\mu_\mu, c\sigma_\mu}(\mu_R) \, 
\chi_{C}(\mu_L,\mu_R) \; / \;d_* 
\end{equation}
and $d_*$ is the normalizing constant. 
The determination of $\chi_{C}(\mu_L,\mu_R)$ is discussed in Section~\ref{sec:constraints};
it is the set $(\mu_L,\mu_R,\mu_{same})$ which results in a monotonic function.
Note that $C$ is of the form $C = \{(\mu_L,\mu_R): a \le \mu_L \le \mu_R \le b \}$ with $a, b$ 
(possibly $-\infty$ and/or $\infty$) determined by the conditioning on $T^*$ and $\mu_{same}$.
In particular, note that $C$ depends on $\mu_{same}$ but we have suppressed this in the notation
for the sake of simplicity.

Closed forms for \eqref{integral1} and the norming constant $d_*$ are unavailable.
However, since the integrals are only two-dimensional, it is straighforward to compute them numerically.
To use a very simple approach, we approximate them by summing over a grid of $(\mu_L,\mu_R)$ values.  
We choose a grid of equally spaced $\mu$ values and then let $G$ be the set of 
$(\mu_L,\mu_R)$ where both $\mu_L$ and $\mu_R$ belong to the grid.

Then, our approximate integrals are
\begin{equation}\label{eq:integral2}
\tilde{p}(r_L, r_R \C T^*, \mu_{same})= \sum_{(\mu_L,\mu_R) \in G \cap C} p(r_L \C \mu_L)\,p(r_R  \C \mu_R)\, \tilde{p}(\mu_L, \mu_R \C T^*, \mu_{same}),
\end{equation} 
where
\begin{equation}
\tilde{p}(\mu_L, \mu_R \C T^*, \mu_{same}) = \phi_{\mu_\mu, c\sigma_\mu}(\mu_L) \phi_{\mu_\mu, c\sigma_\mu}(\mu_R)  \; / \; \tilde{d}_*
\end{equation}
with
\begin{equation}\label{eq:integral3}
\tilde{d}_* = \sum_{(\mu_L,\mu_R) \in G \cap C}  \phi_{\mu_\mu, c\sigma_\mu}(\mu_L) \phi_{\mu_\mu, c\sigma_\mu}(\mu_R).
\end{equation} 
Note that we do not include ``$\Delta \mu$'' terms (the difference between adjacent grid values) 
in our integral approximations since they cancel out.

If $T^1 = T^*$ is accepted, the simulation of $(\mu_L,\mu_R)$ proceeds by sampling from the probability distribution over 
$G \cap C$ given by 
\begin{equation}\label{eq:bivdraw}
\tilde{p}(\mu_L, \mu_R \C r_L, r_R, T^*, \mu_{same}) = \frac{p(r_L \C \mu_L)\, p(r_R  \C \mu_R)\, \tilde{p}(\mu_L, \mu_R \C T^*, \mu_{same})}{\tilde{p}(r_L, r_R \C T^*, \mu_{same})}.
\end{equation}
Note that $\tilde{d}_*$ cancels in \eqref{eq:bivdraw} so that we are just renormalizing 
$$
p(r_L \C \mu_L)\, p(r_R  \C \mu_R)\, \phi_{\mu_\mu, c\sigma_\mu}(\mu_L) \phi_{\mu_\mu, c\sigma_\mu}(\mu_R)
$$
to sum to one on $G \cap C$.

For the calculation of
\begin{equation}\label{eq:oneDint}
p(r_0 \C T^0, \mu_{same}) = \int  p(r_0  \C \mu_0) \, p(\mu_0 \C T^0, \mu_{same}) \, d\mu_0
\end{equation}
where
\begin{equation}
p(\mu_0 \C T^0, \mu_{same}) = \phi_{\mu_\mu, c\sigma_\mu}(\mu_0) \chi_{C} (\mu_0) \; / \; d_0
\end{equation}
and $d_0$ is the normalizing constant 
with the constraint set of the form $C = \{(\mu_0): a \le \mu_0 \le b \}$, 
similar griding can be done to obtain a discrete approximation 
$\tilde{d}_0$ of $d_0$ and a
constrained posterior sample of $\mu_0$.  
Again, $C$ implicitly depends on $T^0$ and $\mu_{same}$.
The grid here would be just one-dimensional. 

Computations for the reverse death move would proceed similarly.
Local moves for $T^0\rightarrow T^*$ beyond birth and death moves may also be similarly applied, 
as long as $\mu_{old}$ and $\mu_{new}$ are each at most two dimensional  
since beyond two dimensions, grids become computationally demanding.
For example, $T^0\rightarrow T^*$  obtained by changing a splitting rule 
whose children are terminal nodes would fall into this category.
In all our examples, we use birth/death moves and draws of a 
single $\mu$ component given $T$ and all the remaining elements of $M$.

The approach outlined above for birth/death moves involves two bivariate integrals 
and two univariate integrals which we approximate with two sums over a bivariate grid
and two sums over a univariate grid.
In practice, we reduce the computational burden
by letting 
$\tilde{d}_*$ and $\tilde{d_0}$ 
equal one and then compensating for this omission with an adjustment of our $T$ prior.
For example, in a birth move, setting the $d$'s to one ignores a factor
$\tilde{d}_0 \; / \; \tilde{d}_*$ in our ratio.
Note that from \eqref{eq:dstar} $d^*$ is just the constrained integral of the product of two univariate normal densities.
Without the constraint, the integral would be one. The more our monotonicity constraint limits the
integral (through $\chi_C(\mu_L,\mu_R)$), the smaller $d^*$ is.
Similary, $d_0$ is a constrained univariate integral.
However, in a birth step, 
$d_*$ is typically more constrained than $d_0$.
Hence, $\tilde{d}_0 \; / \; \tilde{d}_*$ is a ratio depending on $T^0$ and $T^*$ which we expect to be greater than one.
Note that $d_0$ only depends on $T^0$ and $d_*$ only depends on $T^*$ (that is, not on $\mu_{same}$).

We compensate for the omission of $\tilde{d}_*$ and $\tilde{d_0}$ by letting 
$\alpha = .25$ and $\beta = .8$ 
rather than using
standard BART default values of 
$\alpha = .95$ and $\beta = 2$. 
With $\alpha = .25$ and $\beta = .8$, $p(T^*)/p(T^0)$ is larger mimicking the effect of
the omitted $d$ ratio.  We have found that with these choices we get tree sizes comparable to those obtained in unconstrained BART.
The values $\alpha = .25$ and $\beta = .8$ are used in all our examples.

{  Finally, we should comment on the additional cost in time introduced by using the numerical approximation with the localized MH algorithm for  constrained predictors,  instead of the usual conjugate MH algorithm for unconstrained predictors.  We have found for example that with a $20 \times 20$ grid size (which yields excellent results for each two-dimensional numerical approximation), the time per MCMC iteration is about 5 times slower for each constrained predictor as compared to the time per iteration for each unconstrained predictor.  Note that with only a small number of constrained predictors in mixed monotonicity settings, this increased burden will be relatively small, and will not affect the speed of handling any other unconstrained predictors under consideration. }

\section{Examples}\label{sec:Examples}

In this section, we illustrate and compare the performance of mBART with related methods on three simulated and two real examples.  For the real examples, where the true regression function is unknown, we include the standard linear model in the comparisons.  In all cases we use default priors for mBART and BART, but remind the reader that for best out-of-sample results, it may be wise to consider the use of cross-validation to tune the prior choice as illustrated in CGM10.

{Throughout the examples, the ``fit'' of BART or mBART at a given $x$ refers to the posterior mean of $f(x)$ estimated by averaging the $f$ draws evaluated at $x$. The $95\%$ credible intervals used to gauge the posterior uncertainty 
about $f(x)$ are obtained simply as the intervals between the upper and lower $2.5\%$ quantiles of these $f$ draws at $x$.  Just as for BART, the uncertainty intervals for mBART will be seen to behave sensibly, for example, by widening where there is less data and for $x$ values far from the data.}

\subsection{The Smoothing Effectiveness of mBART}\label{subsec:smooth}

\begin{figure}
\centerline{\includegraphics[scale= .35]{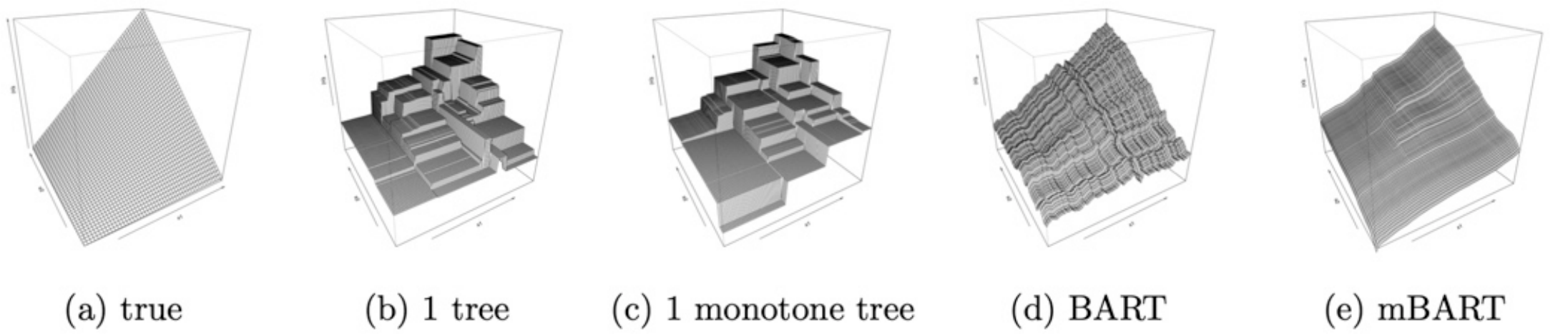}}
\caption{\small Estimating $f(x_1,x_2)$.  From left to right:  the true $f$, a single tree model, a monotone single tree model, BART and mBART.}
\label{figsmth}
\end{figure}

We begin with a visual illustration of the performance of mBART relative to simpler Bayesian tree model approaches on $n=200$ independent simulated observations from the simple two-dimensional predictor model
\begin{equation}
Y = x_1 \, x_2 + \epsilon, \;\; \epsilon \sim N(0, \sigma^2)
\end{equation}
where $x_1,x_2$ $\sim$ Uniform(0,1).  The mean function  $f(x_1,x_2) = x_1 \, x_2$, displayed in Figure \ref{figsmth}a, is smoothly monotone over the (0,1) range of the $x$'s. 

The remaining plots in Figure \ref{figsmth} display successive estimates of the $f$ surface obtained by a single tree model (Bayesian CART), a monotone constrained single tree model, BART and mBART.   The fit of the single tree model in Figure \ref{figsmth}b is reasonable, but there are aspects of the fit which violate monotonicity.  The fit of the single monotone constrained tree in  Figure \ref{figsmth}c is better and more representative of the true $f$. The unconstrained ordinary BART fit in Figure \ref{figsmth}d is much better, but not monotone.  Finally, the correctly constrained mBART fit in Figure \ref{figsmth}e is much smoother, a noticeable improvement over all.  These comparisons highlight the smoothing effect of both summing many trees and of constraining them to be monotone.  This same effect would of course occur in higher dimensions with many $x$'s, but would not allow for such a simple revealing visual illustration. 



\subsection{Comparing Fits and Credible Regions of BART and mBART}\label{subsec:onedsim}

\begin{figure}[htp]
\centerline{\includegraphics[scale=.35]{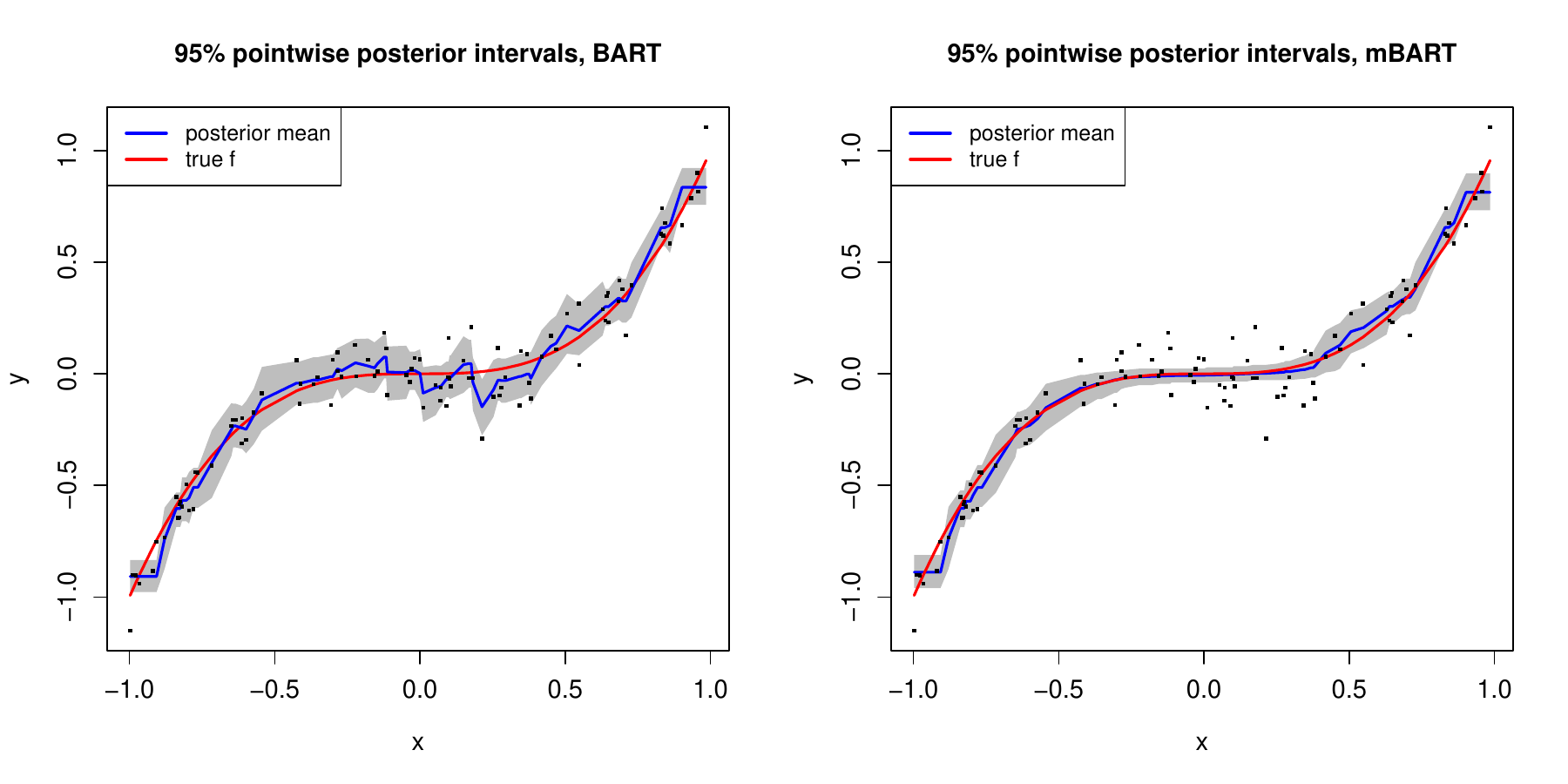}}
\vspace{-.5cm}
\caption{\em \small
Comparing BART and mBART inferences for a monotone one-dimensional example $f(x) = x^3$.   The mBART fits are better throughout and the 95\% pointwise intervals for $f(x)$ are tighter.
}
\label{fig:1d_interval-comparison_bart-vs-mbart}
\end{figure}

{  To facilitate simple visual comparisons of the fits and credible regions obtained by BART and mBART, we continue with a one-dimensional example.  For this purpose, we simulated  200 replications of $n = 100$ independent observations from the single predictor, monotone increasing model
\begin{equation}\label{eq:x3}
Y = x^3 + \epsilon, \;\;\;\; \epsilon \sim N(0,\sigma^2),
\end{equation}
with $\sigma = 0.1$ at $x$ values uniformly sampled from [-1,1].

For a typical one of these data sets, Figure~\ref{fig:1d_interval-comparison_bart-vs-mbart} displays the fits and 95\% pointwise credible intervals for BART on the left and mBART based on a nondecreasing monotonicity assumption on the right.  The improvement of mBART over BART is immediately apparent.  mBART is far smoother and faithful to $f$, with tighter credibility intervals reflecting a reduction of uncertainty, even more so nearer the center of the data where $f$ is flatter.    Adding only the prior information that $f$ is monotone increasing appears to have substantially improved inference.}

\begin{figure}[htp]
\centerline{\includegraphics[scale=.35]{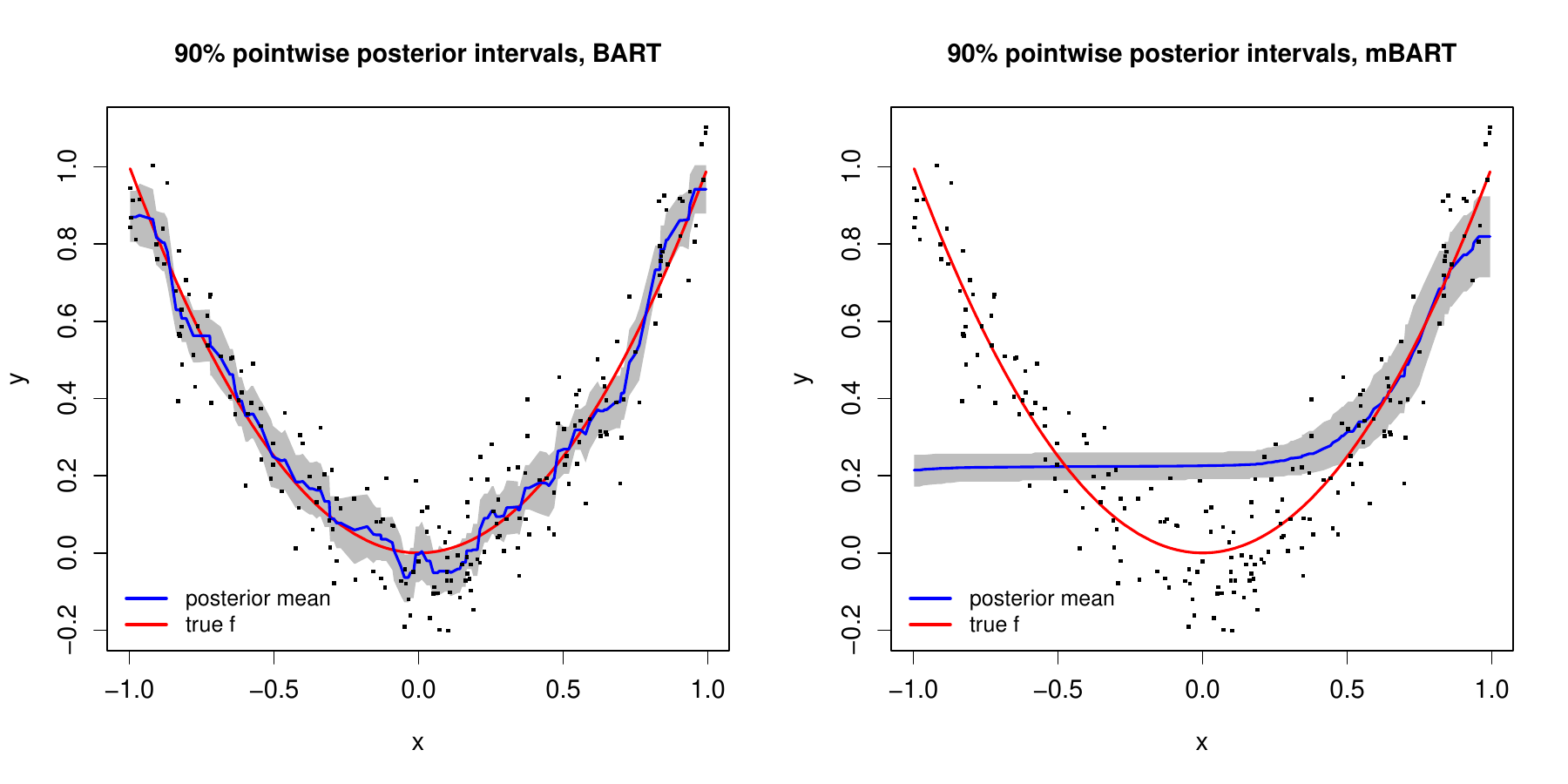}}
\vspace{-.5cm}
\caption{\em \small
Comparing BART and mBART inferences for a non-monotone one-dimensional example $f(x) = x^2$.  Comparison reveals the obvious lack of monotonicity of the function.
}
\label{fig:1d_interval-comparison_bart-vs-mbart_x2}
\end{figure}

{  Supporting the persistence of the mBART improvements seen in Figure~\ref{fig:1d_interval-comparison_bart-vs-mbart}, the average in-sample root mean square error of the mBART fits was 34.8\%  smaller than that of the BART fits, and the average width of the mBART 95\% credible intervals was 40.3\% smaller than the mBART intervals, over all 200 data replications.   Furthermore, the average coverage of $f(x)$ was 94.1\% for the mBART intervals and 96.1\% by the BART intervals, supporting the practical reliability of these intervals in terms of their frequentist calibration.

We should emphasize here that despite the potential improvements that mBART offers, the validity of mBART inferences in practice will rest on the validity of the monotonicity assumptions,  which themselves would presumably be based on compelling subject matter considerations.  A data analyst who observed only the data in Figure 6, could at best conclude from the comparison of BART with mBART that an assumption of monotonicity was plausible.  Fortunately, in settings with more pronounced violations of monotonicity,  comparisons of BART with mBART can readily reveal that mBART should be avoided.   To illustrate this point, we simulated another  $n = 100$ independent observations as above, but this time with an underlying quadratic function  $Y = x^2 + \epsilon$.  For this data, Figure~\ref{fig:1d_interval-comparison_bart-vs-mbart_x2} displays the fits and 95\% pointwise credible intervals for BART on the left and mBART based on a nondecreasing monotonicity assumption on the right.  Comparison of BART and mBART here clearly reveals the implausibility of the monotonicity assumption and the obvious superiority of BART.  Such comparisons can be made in higher dimensional settings with  versions of the conditional view plots used in Sections \ref{subsec:cars} and \ref{subsec:fin}. } 

\subsection{Improving the RMSE with Monotone Regularization}\label{subsec:oos-5d}






We next turn to a comparison of the out-of-sample predictive performance of BART and mBART for data simulated from the five predictor model
$$
Y = x_{1} x_{2}^2 + x_{3} x_{4}^3 + x_{5} + \epsilon, \;\;\;\; \epsilon \sim N(0,\sigma^2),
$$ 
where $x_1,\ldots, x_5$ iid $\sim$ Uniform(0,1).  The mean function here $f(x) = x_{1} x_{2}^2 + x_{3} x_{4}^3 + x_{5}$ is monotonic over the (0,1) range of all the components of $x = (x_1,\ldots, x_5)$.

{  For this setup, we replicated data for five values of the error standard deviation, $\sigma$ = 0.2, 0.5, 0.7, 1.0, 2.0, to explore how rapidly the predictive performance of BART and mBART would degrade as the signal-to-noise ratio decreased.  As we will see, for small $\sigma$, there is little difference in the performance as BART is able to infer the function with very little error.  However, as $\sigma$ increases, the additional information that the function is monotonic becomes more and more useful as mBART outperforms BART by larger and larger amounts.

%
%

For each value of $\sigma$ we simulated 200 data sets,
each with 500 in-sample (training) observations and 1,000 out-of-sample (test) observations.
For the training data, we drew $x$ and $y$, while for the test data we only drew $x$.
For each simulated data set,  we computed the BART to MBART ratio of their out-of-sample RMSE = $\sqrt{\frac{1}{1000} \sum_{i=1}^{1000} (f(x_i)-\hat{f}(x_i))^2}$ estimates, 
where $f$ is the true function, $\hat{f}(x_i)$ is the posterior mean, and the $x_i$
are the test $x$  vectors. }

\begin{figure}[htp]
\centerline{\includegraphics[scale=.36]{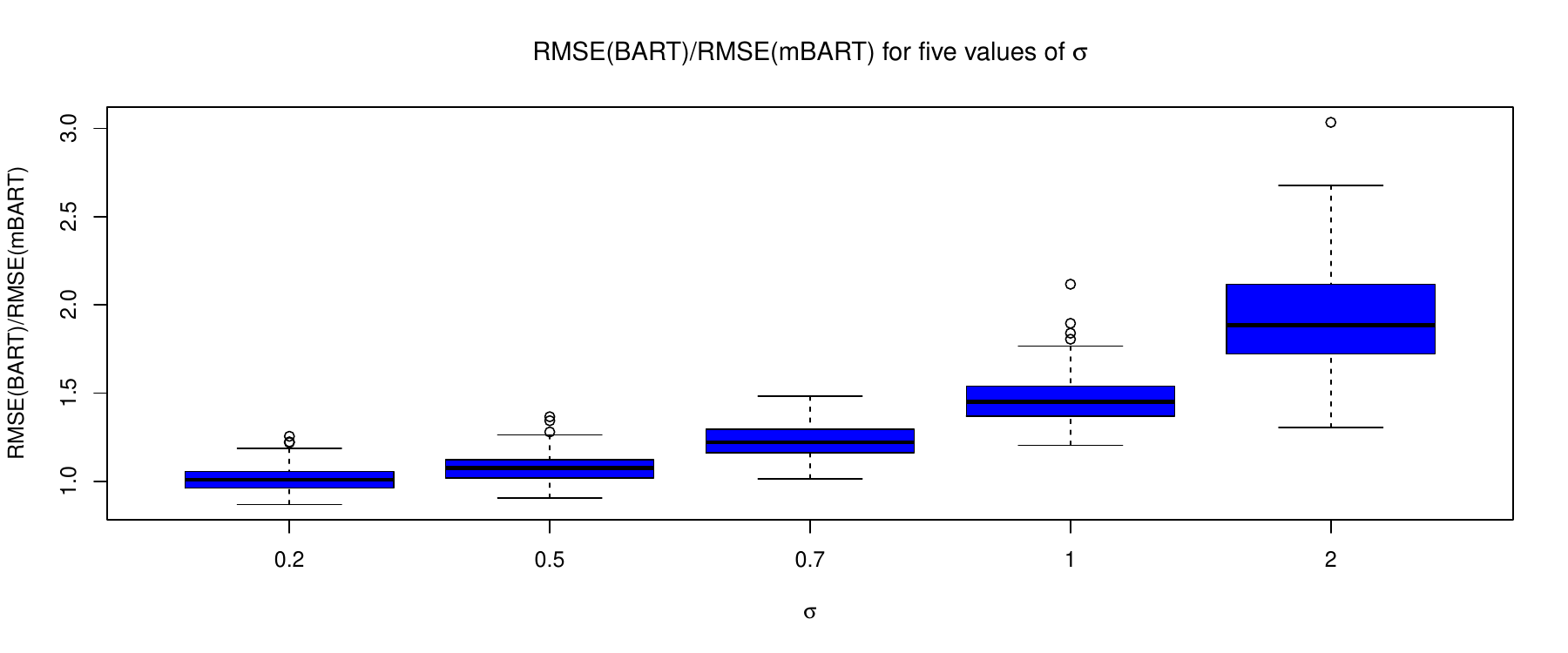}}
\vspace{-.5cm}
\caption{\em \small
Out-of-sample RMSE ratio comparisons of BART and mBART.
}
\label{fig:5-d_sim}
\end{figure}

{  These RMSE ratio results are displayed in Figure~\ref{fig:5-d_sim}.
Each boxplot depicts the 200 RMSE ratio values at each of the five levels of $\sigma$. For the smallest value of $\sigma$ both methods give similar results.
But as $\sigma$ increases, mBART increasingly outperforms BART by greater amounts.   Intuitively, the monotone constraint encourages mBART to disregard variation which runs counter to the prescribed monotonicity.  This shape constrained regularization leads to improved predictions and guards against overfitting irrelevant variation when the monotonicity constraints are justified.  This improvement becomes more and more pronounced as the signal-to-noise ratio decreases. Indeed, the average of the RMSE ratios was 1.01, 1.08, 1.23, 1.47, 1.93  at $\sigma$ = 0.2, 0.5, 0.7, 1.0, 2.0, respectively.}

{  At each value of $\sigma$, we also evaluated the out-of-sample performance of the 95\% credible intervals obtained by BART and mBART on each our 200 simulated data sets.   Table \ref{tab:wc} reports the average interval width and frequentist coverage of $f(x)$ over the 1,000 out-of-sample values.  The average width of the mBART intervals is dramatical smaller, increasingly so as $\sigma$ increases.  The average coverage of the BART intervals at around 99\% is higher than the 95\% credibility levels, whereas the coverage of the mBART intervals increases from 90.1\% to 97.9\% as $\sigma$ increases.  For practical purposes, these calibrations support good reliability, especially when $\sigma$ is larger and the mBART improvement is most valuable.   
}

\vspace{.1cm}

\begin{table}
\begin{center}
{\small
 \begin{tabular}{||c | c c c c c ||} 
 \hline
$\sigma$ & 0.2 & 0.5 & 0.7 & 1.0 & 2.0 \\ 
 \hline\hline
Width BART & 0.44 & 0.81 & 1.06 & 1.46 & 2.79 \\ 
 \hline
Width mBART & 0.27 & 0.46 & 0.56 & 0.71 & 1.17 \\
 \hline \hline
Coverage BART & 98.8\% & 99.1\% & 99.3\% & 99.4\% & 99.6\%  \\
 \hline
Coverage mBART  & 90.1\% & 90.3\% & 92.8\% & 95.7\% & 97.9\%  \\  
 \hline
\end{tabular}
}
\end{center}
\vspace{-.3cm}
\caption{\em \small
Average width and coverage of BART and mBART 95\% credible intervals.
}
\label{tab:wc}
\end{table}

\subsection{Used Car Prices}\label{subsec:cars}

For our first {real} example, our data consists of 1,000 observations and $y$ is the sale price of a used Mercedes car.  
Our explanatory $x$ variables are: 
(i) the mileage on the car ({\tt mileage}), 
(ii) the year of the car ({\tt year})
(iii) feature count ({\tt featureCount}) and 
(iv) has the car had just one owner (1 if yes, 0 if no) ({\tt isOneOwner}).
{ Conditionally on the other variables, we assumed that on average, 
a car with higher mileage would sell for less, a newer car would sell for more, a car with just one owner would sell for more, and a car with higher feature count would sell for less}.  To conveniently characterize all of these expected relationships as monotone increasing, we multiplied {\tt mileage} and  {\tt featureCount} by -1.

Before proceeding with mBART, we ran an ordinary multiple linear regression which produced the following output.  We see that all the signs are positive and {\tt featureCount} is ``significant''. {   It turns out we misunderstood the nature of this variable as will be discussed. Nevertheless, we left {\tt featureCount} in the presented analysis as 
``adding a variable by accident'' is a realistic possibility, and one which mBART 
turns out to handle nicely.}

{\scriptsize
\begin{verbatim}
Coefficients:
               Estimate Std. Error t value Pr(>|t|)    
(Intercept)  -5.427e+06  1.732e+05 -31.334  < 2e-16 ***
mileage       1.529e-01  8.353e-03  18.301  < 2e-16 ***
year          2.726e+03  8.613e+01  31.648  < 2e-16 ***
featureCount  3.263e+01  9.751e+00   3.346 0.000851 ***
isOneOwner    1.324e+03  6.761e+02   1.959 0.050442 .  
---
Signif. codes:  0 ‘***’ 0.001 ‘**’ 0.01 ‘*’ 0.05 ‘.’ 0.1 ‘ ’ 1

Residual standard error: 7492 on 995 degrees of freedom
Multiple R-squared:  0.8351,	Adjusted R-squared:  0.8344 
F-statistic:  1260 on 4 and 995 DF,  p-value: < 2.2e-16
\end{verbatim}
}

\begin{figure}[htp]
\centerline{\includegraphics[scale=.25]{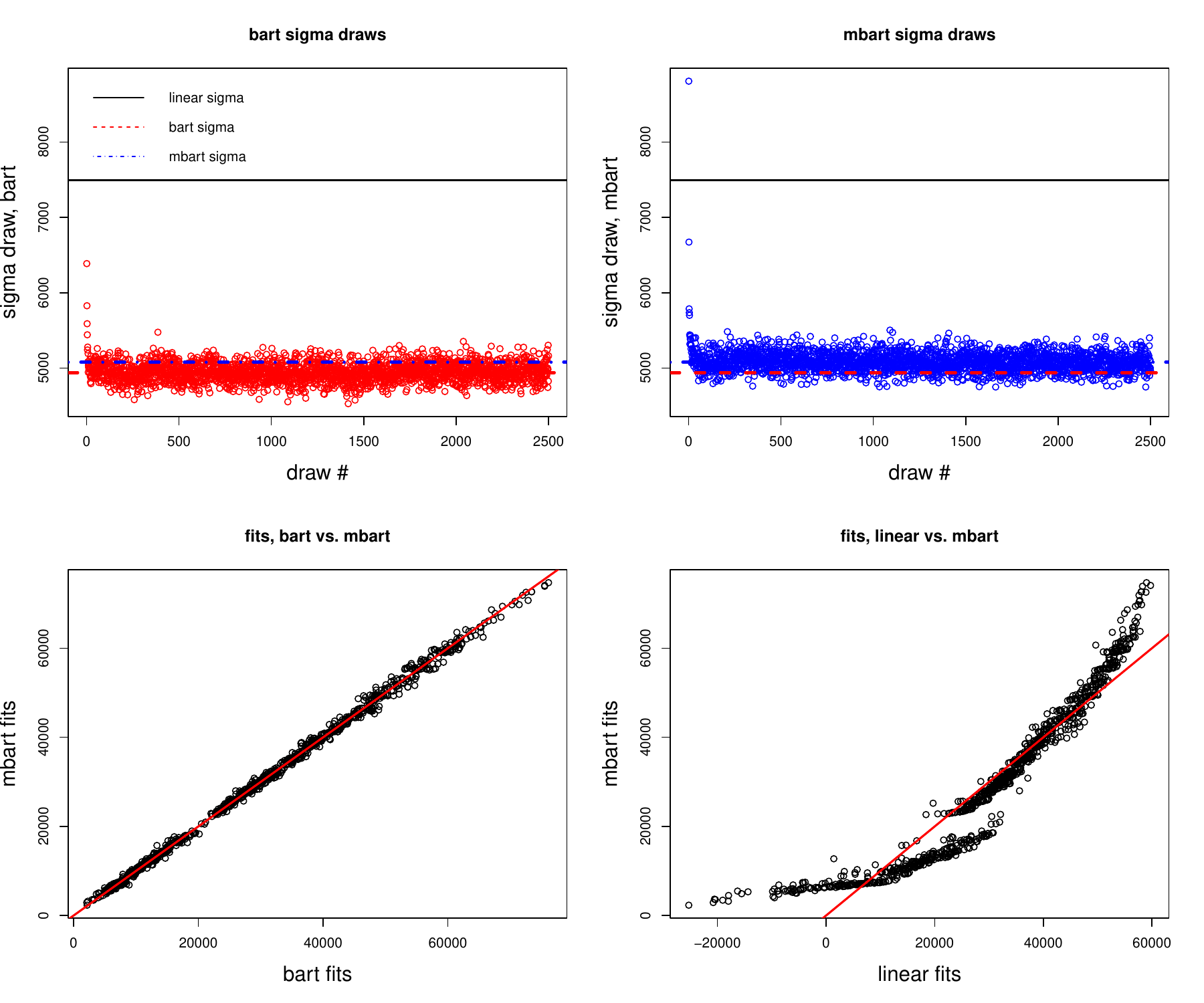}}
\vspace{-.5cm}
\caption{\em \small
Car price example.
Top row: $\sigma$ draws from BART (left panel), $\sigma$ draws  from mBART (right panel).
Solid horizontal line at least squares estimate of $\sigma$.
Bottom row:
BART versus mBART (left panel), and linear fits versus mBART (right panel).
}
\label{fig:cars-bart-vs-mbart-linear}
\end{figure}

{ Figure~\ref{fig:cars-bart-vs-mbart-linear} displays aspects of the inference from the linear model, BART, and mBART.
The top left panel plots the BART MCMC draws of $\sigma$, the top right panel plots the mBART MCMC draws of $\sigma$, and in each plot the estimate of $\sigma$ from the linear regression is indicated by a horizontal solid line.
Both BART and mBART quickly burn-in to $\sigma$ values much smaller than the
least squares estimate indicating much tighter fits.  The monotonicity constraint renders
slightly larger $\sigma$ draws.
The bottom left panel, which plots the BART fits versus the mBART fits, shows them to be quite similar.
In contrast, the bottom right panel, which plots the linear fits versus the mBART fits, shows clear differences between the two.}

\begin{figure}[htp]
\centerline{\includegraphics[scale=.42]{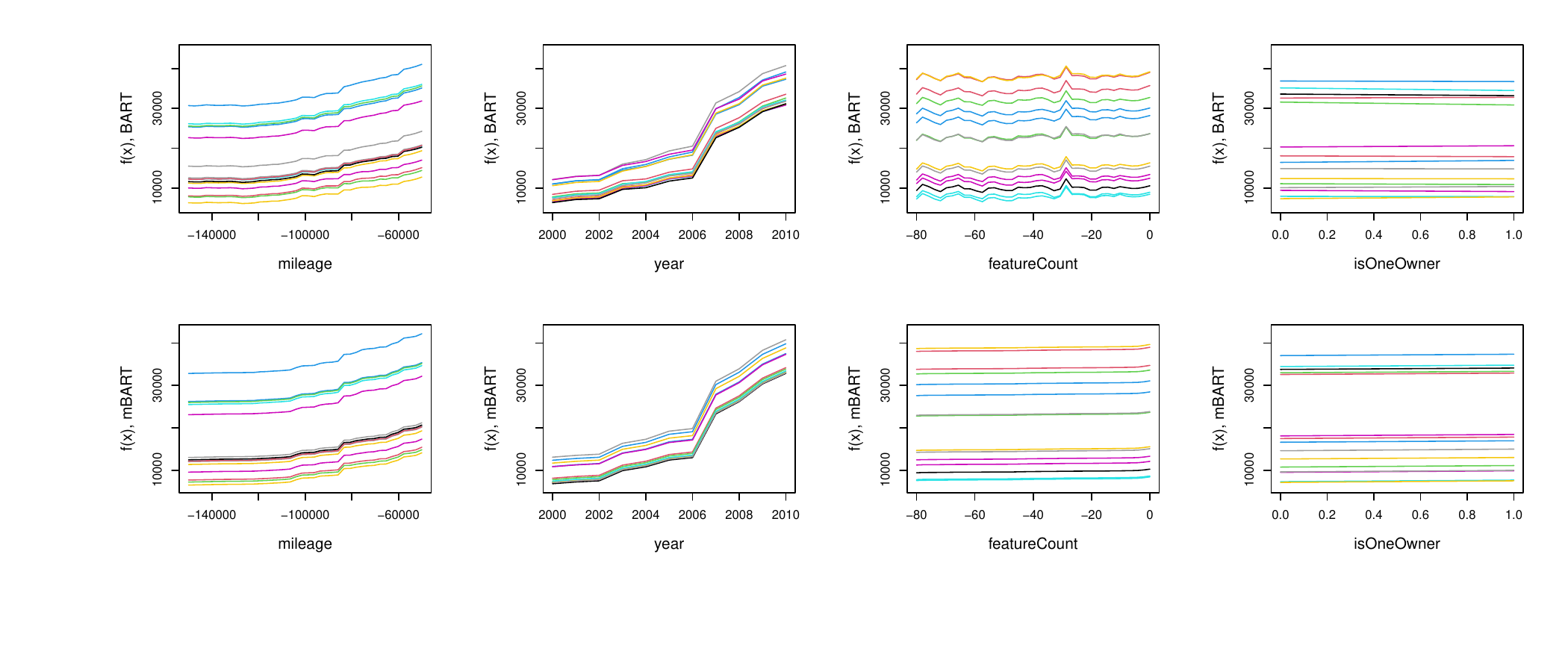}}
\vspace{-1cm}
\caption{\em \small
Car price example.
Conditional effects of mileage, year, featureCount and isOneOwner, in rows from left to right,
for BART (top panels) and mBART (bottom panels).
}
\label{fig:cars_marginal-effects-1234}
\end{figure}

{Figure \ref{fig:cars_marginal-effects-1234} displays the estimated conditional effects for the four variables in $x$,  {\tt mileage}, {\tt year}, {\tt featureCount} and {\tt isOneOwner}.   To visualize the conditional effects from the BART/mBART fits we construct $x$ vectors such that
the $x$ coordinate of interest varies while the others are held fixed.
In each panel, we see the estimate of $f(x)$ with the year values of $x$ indicated on the horizontal axis.  The various curves in the figure  correspond to different fixed levels of the other three variables
in $x$. We picked a grid of values for each variable and then
constructed a design matrix composed of all possible combinations.  
To keep the plots readable, we conditioned on a random sample of these value combinations, and held them fixed as we varied the variable of interest, so that not all possible curves are plotted in each panel.}

Although the conditional effects of {\tt mileage} and {\tt year} in 
Figure~\ref{fig:cars_marginal-effects-1234}
are similar for BART and mBART, 
the mBART fits are smoother and everywhere monotonic,
while the BART fits exhibit slight dips.  {Fundamental is the observed monotonicity of all the conditional effects,
 reflecting mBART's ability to impose monotonicity in a multivariate setting.} 
For {\tt featureCount}, the difference between the BART and mBART conditional effect plots is quite striking.
The monotonic constraint forces a flat lining of the mBART estimates, dramatically indicating
the absence of an effect, in sharp contrast to the very significant t-value of 3.346 in the {\tt R} multiple regression output.
After obtaining these results,
we checked back with the source of the data and found that we had misunderstood the 
variable {\tt featureCount} and in fact, there was no reason to expect it to be predictive
of the car prices!  It measured web activity of a shopper and not features of the actual car.  { Together, the plots in   Figure~\ref{fig:cars_marginal-effects-1234} indicate that from a practical standpoint, only {\tt mileage} and {\tt year} matter as price predictors here.}

\begin{figure}[htp]
\vspace{-.1cm}
\centerline{\includegraphics[scale=.23]{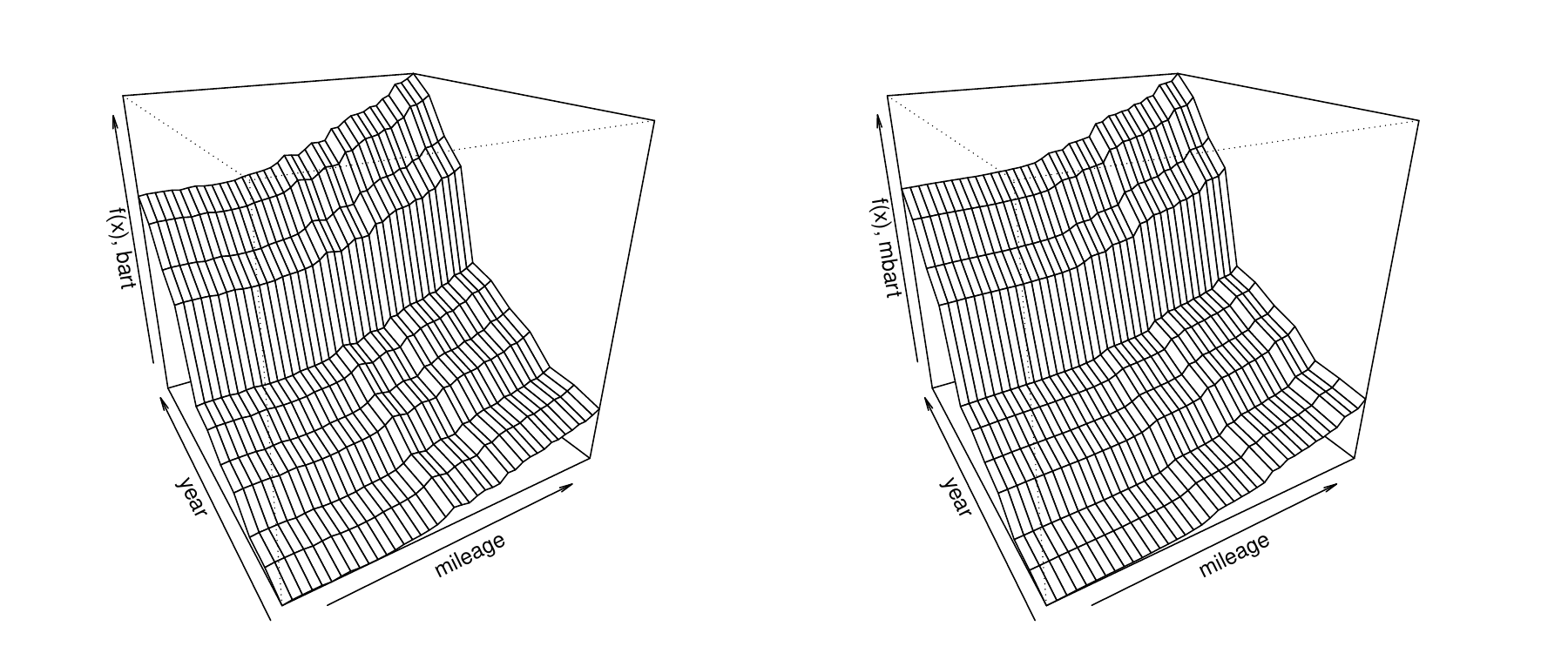}}
\vspace{-.3cm}
\caption{\em \small
Car price example.
Bivariate plot of fitted price vs mileage and year.
BART (left panel), mBART (right panel).
{ Recall that mileage and year have been multiplied by -1, so the fitted price surface is descreasing as actual mileage and actual year are increased.}}
\label{fig:cars_persp_mileage-year}
\end{figure}

Figure~\ref{fig:cars_persp_mileage-year} plots the bivariate fitted surface
for expected price as a function of {\tt mileage} and {\tt year} for fixed values 
of {\tt featureCount} and {\tt isOneOwner}.
The BART fit is on the left and the mBART fit is on the right.
While similar, the mBART fit is smoother, {  and far more appealing.
When presenting results to non-statisticians, the implausible
non-monotonic behavior can be very confusing.}

\begin{figure}[htp]
\centerline{\includegraphics[scale=.23]{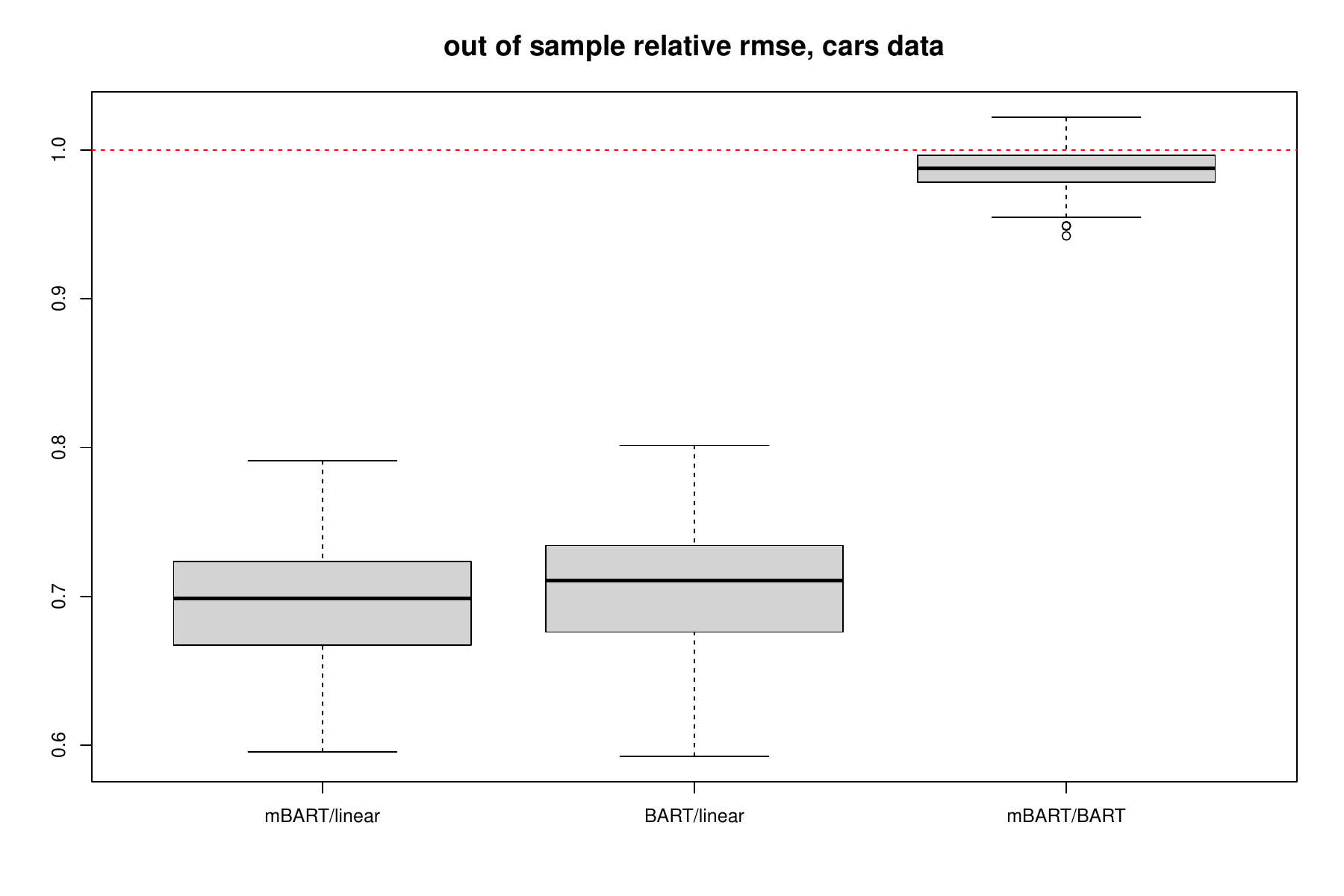}}
\vspace{-.5cm}
\caption{\em \small
Car price example.
Out-of-sample RMSE ratios. 
mBART  and BART are superior to linear, while mBART is modestly better than BART.}
\label{fig:cars-oosrp}
\end{figure}

We conducted a simple out-of-sample experiment to check for over-fitting:
200 times we randomly selected 75\% of the data to be in-sample and predicted
the remaining 25\% of the $y$ values given their $x$ values using linear regression,
BART, and mBART.  {  For each repetition, we evaluated the RMSE ratios of   mBART to linear, and of mBART to BART.  Boxplots of these ratios for the 200 repetitions are displayed in
Figure~\ref{fig:cars-oosrp}.}
Both mBART and BART are dramatically better than the linear predictions, while
mBART provides a more modest improvement over BART.

\subsection{The Stock Returns Example}\label{subsec:fin}

An important and heavily studied problem in Finance is the predictability of stock market returns.
Can we measure characteristics of a firm ($x$) that can help us predict a future return ($y$)?
The data are monthly and the $x$'s are measured the previous month so that the relationship
being studied is predictive.  
Our $y$ is actually {\it excess} return, the difference between the return for a firm 
and the average return for firms that month.  While still very useful in practice, 
predicting the excess return is easier than predicting the whole return.

Often in predictability studies, predictive models are fit for each month and then
rolling windows of months are considered.  { For this example, we focus on fitting models for the returns of $n$ = 1,531 firms for the single month December 1981, (picked randomly from a much larger data set of 594 months), to compare mBART with BART and linear regression.  Since the modeling is done for each month, it makes sense to focus on a particular month to see how different approaches might work. Note that the predictive models uncovered here are descriptive rather than reflective
of the actual averaging rolling fit predictive mechanism used in practice}.

We used four predictive variables in $x$. 
{\tt logme}: market equity (logged), 
{\tt r1}: previous return,  
{\tt gpat}: gross profitability ((sales minus cost of goods sold) / total assets), 
and {\tt logag}: growth in total assets (logged).
Remember, $x$ is lagged.  
Although log transformations of predictors {are} unnecessary for BART and mBART,
these transformations facilitate comparisons with linear regression.

For convenience, we multiplied {\tt logme}, {\tt r1} and {\tt logag} by -1 to obtain 
all monotone increasing relationships in the following multiple linear regression: 

{\scriptsize
\begin{verbatim}
            Estimate Std. Error t value Pr(>|t|)    
(Intercept) 0.028895   0.010052   2.875  0.00410 ** 
logme       0.004461   0.001626   2.744  0.00614 ** 
r1          0.063310   0.020397   3.104  0.00195 ** 
gpat        0.035634   0.007428   4.797 1.77e-06 ***
logag       0.080160   0.010715   7.481 1.24e-13 ***
---
Signif. codes:  0 ‘***’ 0.001 ‘**’ 0.01 ‘*’ 0.05 ‘.’ 0.1 ‘ ’ 1

Residual standard error: 0.07285 on 1526 degrees of freedom
Multiple R-squared:  0.05767,	Adjusted R-squared:  0.0552 
F-statistic: 23.35 on 4 and 1526 DF,  p-value: < 2.2e-16
\end{verbatim}
}

{ The monotonicity implications of this regression are supported by intuition and subject matter theory.  First, the monotonicity for logged market equity {\tt logme} is strongly motivated, as it is widely believed that larger firms are less risky and hence generate lower returns.
Although one might think a high previous return {\tt r1} would lead to a high current return (giving a negative sign in the regression since we multiplied by -1), a tendency for ``short term reversals'' has been found in the literature.}
That gross profitability {\tt gpat} should be positively related to returns as
in the regression makes intuitive sense.  Finally, the monotonicity of the logged growth in total assets {\tt logag} effect is less clear and, indeed, the sign of the regression coefficient can vary from month to month.  However, financial theory suggests that if we interpret our $x$'s as representative of underlying { factors, we would still expect the effect to be monotonic across a set of firms within a given month.}

The $R^2$ in the multiple regression is less than 6\%,
indicating a very low signal to noise ratio.
Bias-variance considerations suggest that only the simplest models can be used
to predict since fitting complex models with such a low signal { is prone to overfitting.  This gives us a strong motivation for examining the fit of mBART.
As we will see, mBART allows us to be more flexible than a simple linear
approach without running the overfitting risks associated with an unconstrained fit given the low signal.}

\begin{figure}[htp]
\centerline{\includegraphics[scale=.25]{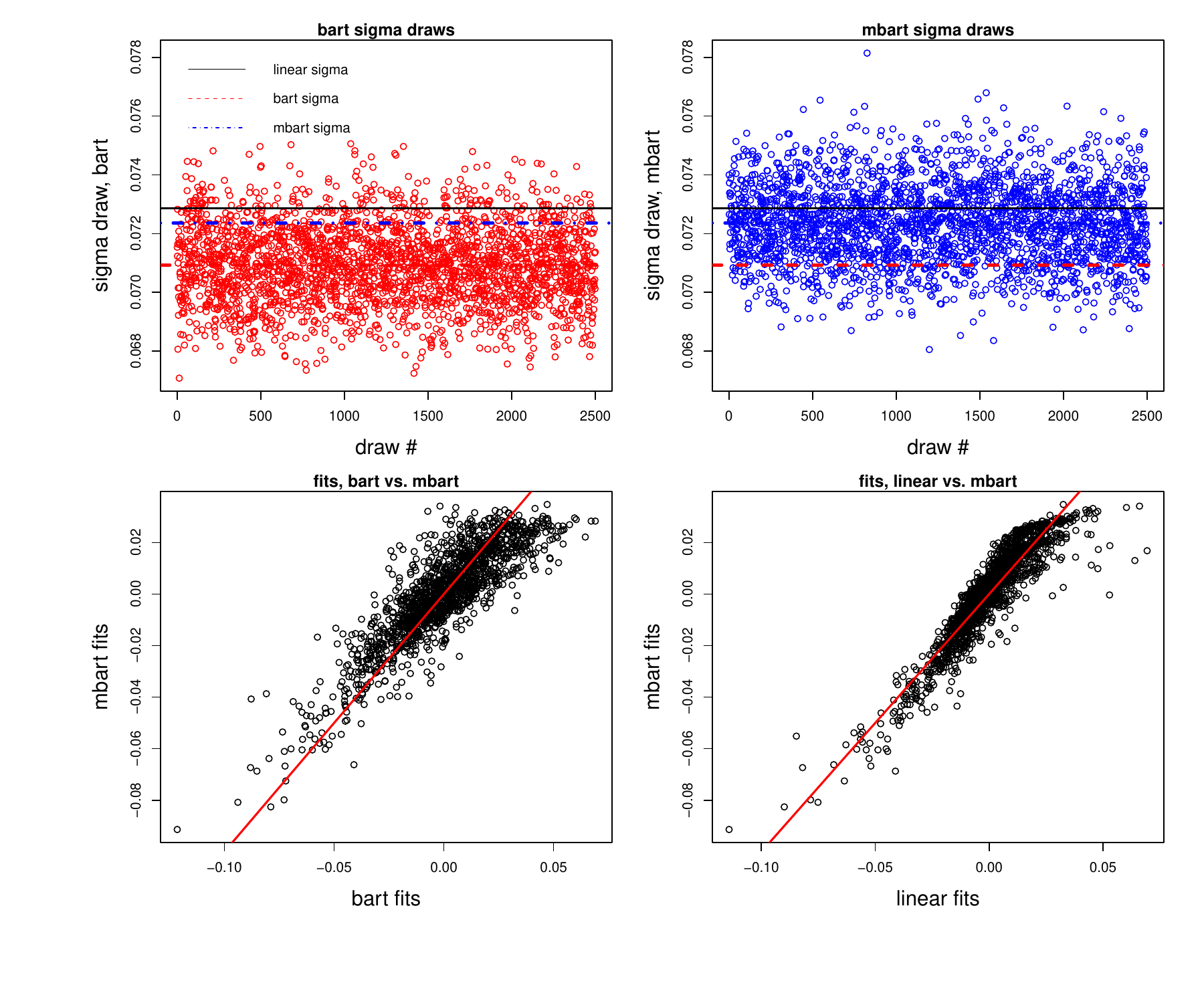}}
\vspace{-.6cm}
\caption{\em \small
Returns example.
Top row: $\sigma$ draws from BART (left panel), $\sigma$ draws  from mBART (right panel).
Solid horizontal line at least squares estimate of $\sigma$.
Bottom row:
BART versus mBART (left panel) and linear fits versus mBART (right panel).
}\label{fig:fin-bart-vs-mbart-linear}
\end{figure}

Figure~\ref{fig:fin-bart-vs-mbart-linear} displays fits from BART, mBART, and a
linear regression (using the same layout as in our previous examples).
The top left plot shows the sequence of $\sigma$ draws from the BART fit,
while the top right plot shows the sequence of $\sigma$ draws from  mBART.
In each plot, a solid horizontal line is drawn 
at the least squares estimate of $\sigma$.
The $\sigma$ draws from the BART fit tend to be smaller than the least squares estimate
while the least squares estimate is right at the center of the mBART fits.
The monotonicity constraint has pulled the BART fit back so that overall, it is more comparable
to the linear fit.  The lower left panel of Figure~\ref{fig:fin-bart-vs-mbart-linear} plots the BART fits
versus the mBART fits and the { lower right} panel plots the linear fits versus
the mBART fits.
Given the very low signal, it is notable that all three methods pick up similar fits.
However, in contrast to { Figure~\ref{fig:cars-bart-vs-mbart-linear}, the mBART fit here appears to be more like the linear fit than the BART fit.}

\begin{figure}[htp]
\vspace{.1cm}
\centerline{\includegraphics[scale=.38]{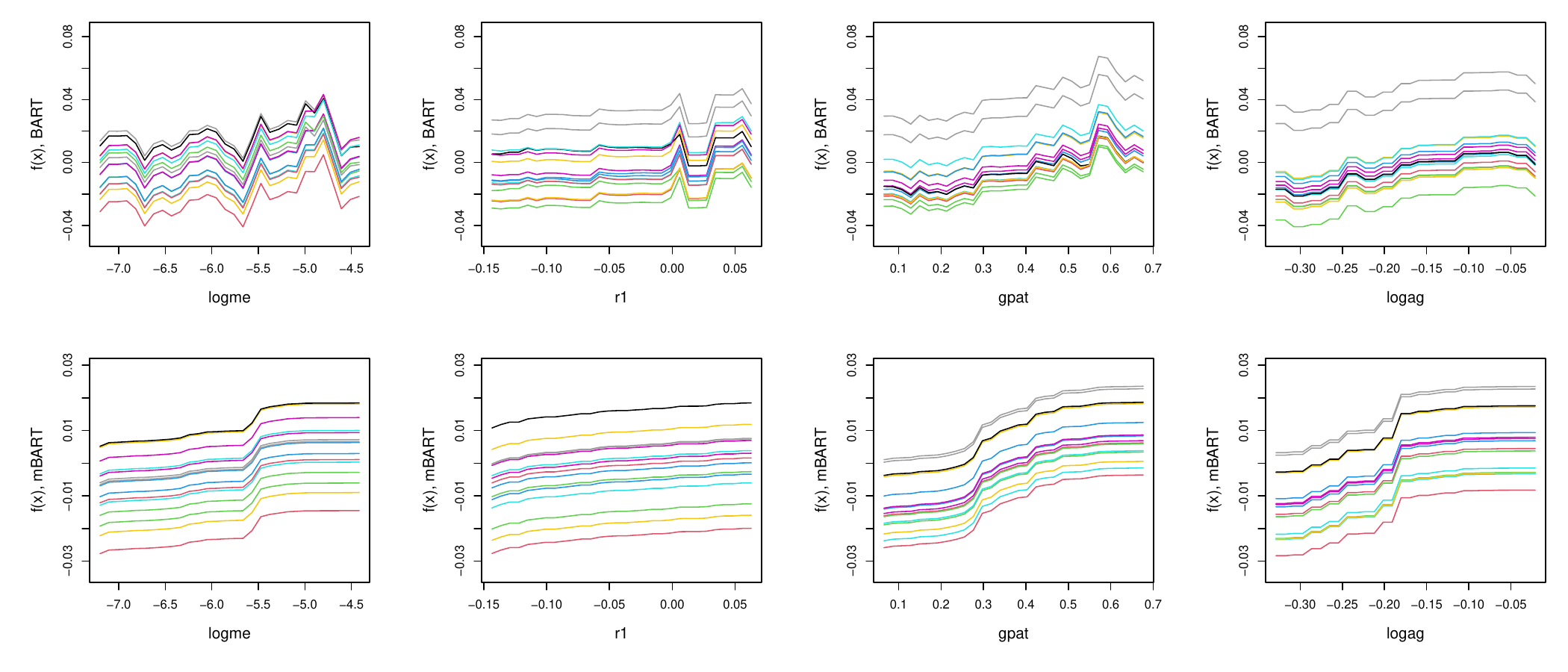}}
\vspace{-.5cm}
\caption{\em \small
Returns example.
Conditional effects of  logme, r1,  gpat and logag,  in rows from left to right,
for BART (top panels) and mBART (bottom panels).
}
\label{fig:fin-effects-1234}
\end{figure}

Figure~\ref{fig:fin-effects-1234} displays the conditional effects using the same
construction and format as in Figure~\ref{fig:cars_marginal-effects-1234} for our car price example.  The contrast between the BART and mBART fits here is quite dramatic.  The mBART fits are much smoother and monotone.
They are  close to linear (especially for {\tt r1}),
but there is an evident suggestion of nonlinearity in places for three of the variables.

\begin{figure}[htp]
\vspace{.1cm}
\centerline{\includegraphics[scale=.25]{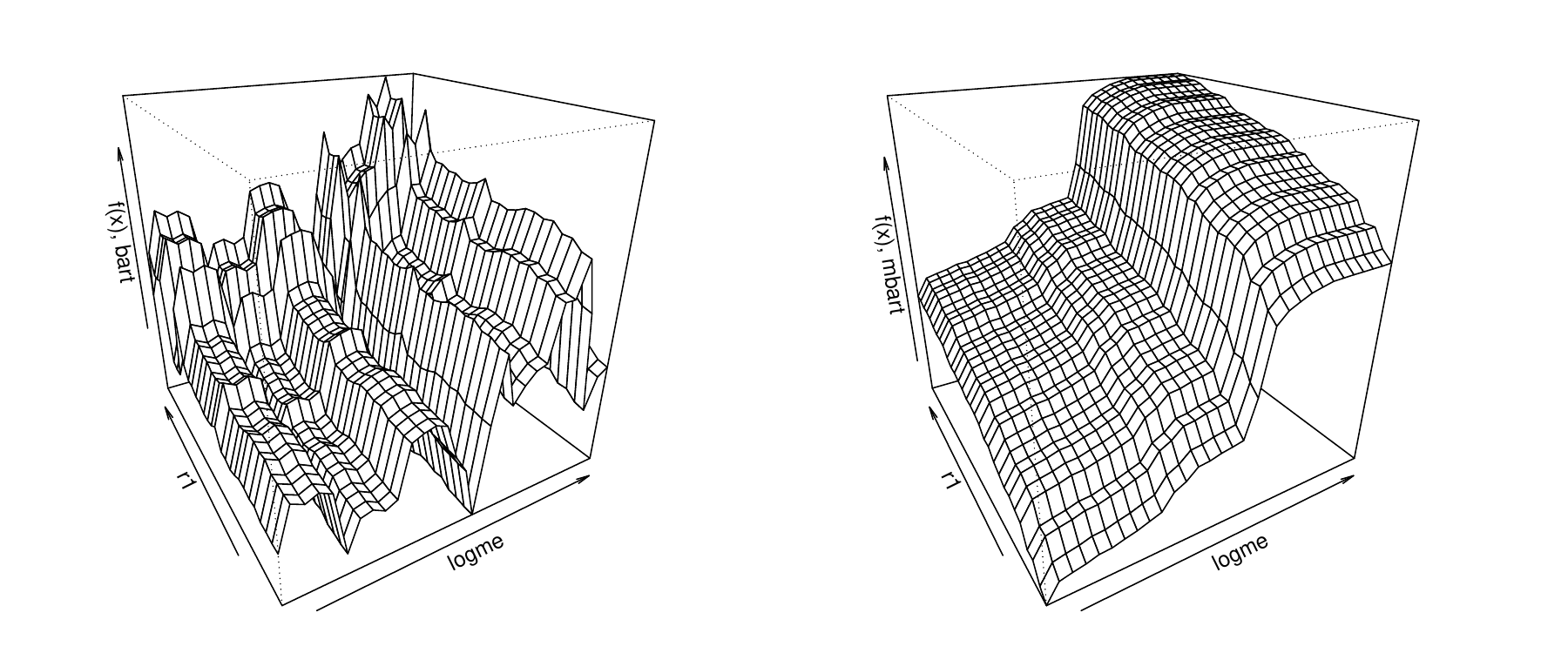}}
\vspace{-.4cm}
\caption{\em \small
Returns example.
Bivariate plot of fitted expected return vs. logme and r1.
BART (left panel), mBART (right panel).
}
\label{fig:persp-fin}
\end{figure}

Figure~\ref{fig:persp-fin} plots the fitted cross section of expected returns
against {\tt r1} and {\tt logme}.
The  unconstrained BART fit seems quite absurd while the mBART fit suggests some
nonlinearity and interaction, but also leaves open the possibility that it is close enough to linear for prediction purposes given the high noise level.

\begin{figure}[htp]
\centerline{\includegraphics[scale=.25]{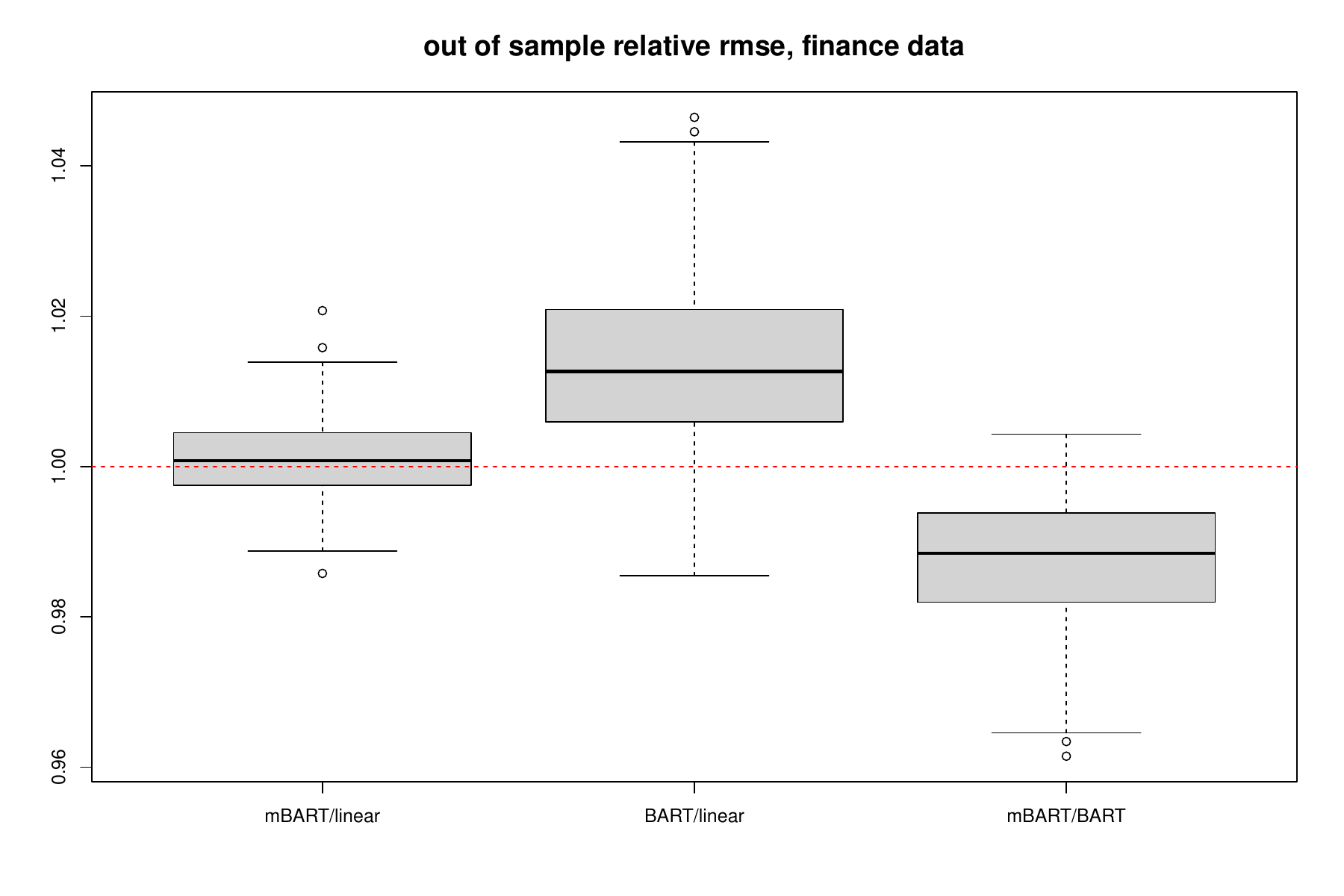}}
\vspace{-.5cm}
\caption{\em \small
Returns example.
Out-of-sample RMSE ratios. 
mBART is comparable to linear, while BART is worse than mBART and linear.
}
\label{fig:finance-oosrp}
\end{figure}

{  To evaluate out-of-sample predictability, we performed a
``stylized'' out-of-sample experiment as
in the previous  used cars example.  That is, we randomly selected 75\% of the data to be in-sample and predicted
the remaining 25\% of the data using linear regression,
BART, and mBART, and repeated this 200 times.
We call this ``stylized'' because 
it is unrealistic to
be interested in using the returns from 75\%
of the firms to predict the rest.
However, this gives a sense for how the procedures work in our particular month.  For each repetition, we evaluated the RMSE ratios of mBART to linear, of BART to linear and of mBART to BART.  Boxplots of these three ratios for the 200 repetitions are displayed in
Figure~\ref{fig:finance-oosrp}}.
{ In contrast to Figure~\ref{fig:cars-oosrp}, we see that here mBART and the linear fit yield very similar results, while BART is now somewhat worse, suggesting a tendency towards overfitting. Given the very low signal-to-noise ratio, the {regularizing} monotonicity constraint of mBART has helped to keep it from fitting variation in the wrong direction.}

{ A key point here is that if you want to consider something more flexible than linear, and interpret the fits on a monthly basis, mBART can give plausible nonlinear results while being predictively equivalent to linear.  At the very least, when monotonicity is a reasonable assumption, we can think of mBART as a convenient ``halfway house'' between the very flexible ensemble method BART and the inflexible linear method.}

\begin{figure}[htp]
\centerline{\includegraphics[scale=.27]{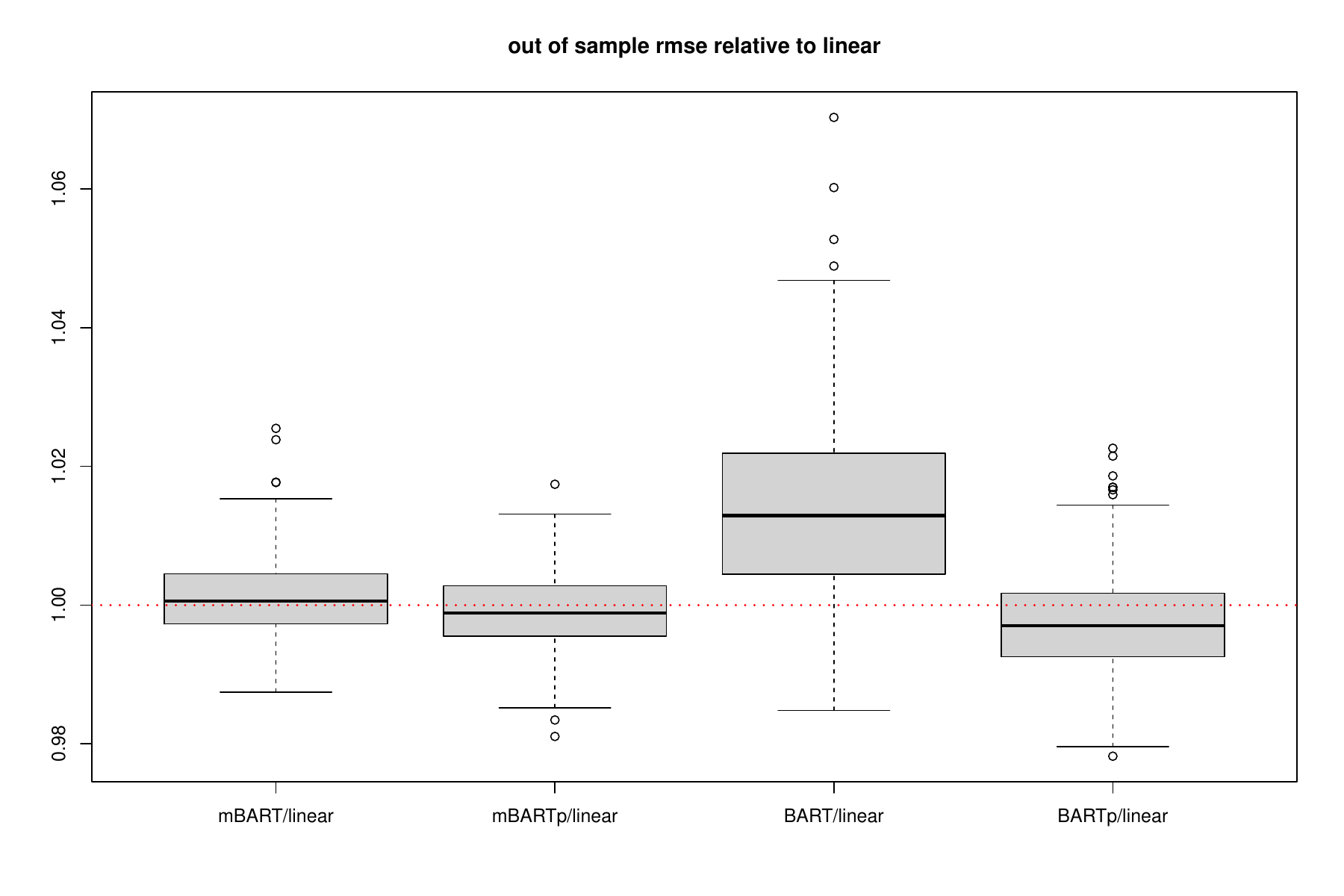}}
\vspace{-.5cm}
\caption{\em \small  Returns example.
RMSE ratio boxplots of  mBART, mBART$_p$, BART and  BART$_p$ all relative to linear.}
\label{relrmse-prior}
\end{figure}

{   Finally, as discussed in Sections \ref{sec:sigmaprior} and  \ref{sec:mprior}, an advantage of the BART and mBART prior specification schemes is their
allowance for subjective calibration.  In the results presented so far
we have used the default, data-based prior which yields
$f(x) \sim N(0,.2^2)$ in the unconstrained case with a corresponding prior 95\% interval of  [-.4,.4].
Here $f(x)$ is the expected return over a single month on a particular firm described by
the attributes in $x$.   

However, it seems implausible that the information in $x$ would suggest an expected return of 40\% in a single month.  Given the weak signal-to-noise ratio in this data, it is likely that the variance of the default prior has been overinflated to cover the range of observed returns, which has been widened by the excessive noise. 
As an alternative we chose the informative prior with $f(x) \sim N(0,.05^2)$ for the unconstrained case, which
gives the 95\% interval [-.1,.1] for the expected return.  While still quite a wide range for the return over a single month,
plus or minus 10\% seems within the realm of plausible predictability.   Note that for the implementation of mBART, we inflate both the default and the informative priors as described in Section 3.3 to account for the monotone constraints. 

It is interesting to compare the predictive performance of mBART and BART under the default priors with their counterparts, denoted mBART$_p$ and BART$_p$, under these subjectively tuned priors.  Figure~\ref{relrmse-prior}  displays RMSE ratio boxplots of  mBART, mBART$_p$, BART and  BART$_p$ all relative to linear, over the 200 repetitions from the Figure~\ref{fig:cars-oosrp} evaluations.  It is interesting that the subjective input has  modestly improved mBART and more substantially improved BART, suggesting that BART flexibility renders it more sensitive to prior calibration.}

\vspace{.2cm}

\section{Discussion}\label{sec:conclusion}

{  In multiple regression problems where the functional form of $E[Y \C x]$  is unknown, subject matter considerations may at least warrant an assumption that $E[Y \C x]$ is monotone in one or more of the predictors in $x$.  mBART is tailor made for such problems.  Inheriting the multidimensional nonparametric modeling flexibility of BART,  mBART can at the same time restrict attention to forms for  $E[Y \C x]$  which are monotonic in any predesignated subset of predictors.  By taking advantage of the additional monotonicity information, this constrained version of BART results in improved estimates and tighter credibility intervals as is illustrated throughout our examples.   These improvements are particularly pronounced in low signal-to-noise contexts.

However, as we emphasized at the end of Section \ref{subsec:onedsim}, these benefits of mBART over BART will rest on the validity of the monotonicity assumptions for which mBART was designed.   When such monotonicity assumptions are in doubt, it will be safer to rely on BART.  However, this raises some interesting directions for further research.  As we also saw in Section \ref{subsec:onedsim}, it will always be useful to compare the output from BART and mBART to judge the plausibility of any monotonicity assumptions.  But even when monotonicity seems plausible, more formal testing procedures such as Bayes factors would be valuable to have.  We plan to report on such developments in future work.

Further important future research directions include the development of theory for mBART.   For example,  in the spirit of Salomond (2014), the added assumption of mononoticity would seem to allow for improved rates of posterior contraction and other refinements of the theoretical results in Rockova and van der Pas (2020), Rockova and Saha (2019) and Linero and Yang (2018) mentioned earlier.  

Finally, it would also be enlightening to investigate empirical and theoretical comparisons of  mBART with the many monotonic alternatives proposed in the references listed in Section \ref{sec:intro}.   Particularly interesting would be the comparison with methods that project unconstrained estimators into monotone spaces, in contrast to mBART which directly constrains the mean regression function to begin with.

Code for mBART is publicly available at:
 https://github.com/remcc/mBART\underline{\hspace{.2cm}}shlib with an R package in the subdirectory mBART.  To install directly in R you can use\\
$>$library(remotes)\\
$>$install\underline{\hspace{.2cm}}github("remcc/mBART\underline{\hspace{.2cm}}shlib/mBART",ref="main")\\
You need to install the R packages remotes and Rcpp.
On  a Mac you  also need to install the Xcode.
On Windows you need to install the Rtools which you can
download from the CRAN R for Windows download page.

\section*{References}
\renewcommand{\baselinestretch}{1.00}

\begin{description} \small

\item {Azzalini, A}. (1985). ``A class of distributions which includes the normal ones.'' {\sl Scand. J. Statist.} {\bf 12} 171-178.

\item {Barlow, R.E.}, {Bartholomew, D.}, {Bremner, J.M.} and {Brunk, H.D.} (1972). {\sl Statistical Inference Under Order Restrictions: Theory and Application of Isotonic Regression.} New York: Wiley.

\item {Bleich, J.}, {Kapelner, A.}, {George, E.I.} and {Jensen, S.T.} (2014).  ``Variable selection for BART: An application to gene regulation.''  {\sl Ann. Appl. Stat.} {\bf 8} 1750-1781.

\item {Cai, B.} and {Dunson, D. B.} (2007).  ``Bayesian multivariate isotonic regression splines: applications to carcinogenicity studies.''  {\sl J. Amer. Statist. Assoc.} {\bf 102} 1158-1171. 

\item {Chen, Y.} and {Samworth, R.J.} (2016). 
 ``Generalized additive and index models with shape constraints.''  {\sl J. R. Statist. Soc. B} \,{\bf 78} 729--754.

\item {Chernozhukov, V.}, {Fernandez-Val, I}. and {Galichon, A.} (2009).  
 ``Improving point and interval estimators of monotone functions by rearrangement.''  {\sl Biometrika} {\bf 96}  559--575.

\item {Chipman, H.}, {George, E.I.} and {McCulloch, R.E.} (1998). Bayesian CART model search (with discussion and a rejoinder by the authors).''  {\sl J. Amer. Statist. Assoc.} {\bf 93} 935-960.

\item {Chipman, H.}, {George, E.I.} and {McCulloch, R.E.} (2010).  ``BART: Bayesian additive regression trees.''  {\sl Ann. Appl. Stat.} {\bf 4} 266-298.

\item {Chipman, H.}, {George, E.I.} and {McCulloch, R.E.} (2013).  ``Bayesian Regression Structure Discovery.'' 
{\sl In Bayesian Theory and Applications}, (Eds, P. Damien, P. Dellaportas, N. Polson, D. Stephens), Oxford University Press, USA.

\item{Hill, J.}, {Linero, A.} and {Murray, J.} (2020).   ``Bayesian Additive Regression Trees: A Review and Look Forward.''  {\sl Annual Review of Statistics and Its Application}  {\bf 7}  251--278.

\item {Holmes, C.C.} and {Heard, N.A.} (2003).  ``Generalized monotonic regression using random change points.''  {\sl Statist. Med.} {\bf 22} 623--638.

\item {Kapelner, A.} and {Bleich, J.} (2016).  ``bartMachine: Machine learning with Bayesian additive regression trees.''  {\sl J. Stat. Softw.} {\bf 70} 1--40.

\item {Kong, M.} and {Eubank, R.L.} (2006).  ``Monotone smoothing with application to dose-response curve.''  {\sl Commun. Statist. B-Simul.} {\bf 35} 991-1004.

\item {Lavine, M.} and {Mockus, A.} (1995).  ``A nonparametric Bayes method for isotonic regression.'' 
{\sl J. Stat. Plan. Inference}. {\bf 46}, 235-248.

\item {Lenk, P.J.} and {Choi, T.}  (2017).    ``Bayesian analysis of shape-restricted functions using Gaussian process priors.''   {\sl Statistica Sinica} {\bf 27} 43-69.

\item {Lin, L.} and {Dunson, D.B.} (2014).  ``Bayesian monotone regression using Gaussian process projection.''  {\sl Biometrika} {\bf 101} 303-317.

\item {Lin, L.}, {St.~Thomas, B.}, {Piegorsch, W.W.}, {Scott, J.} and {Carvalho, C.} (2019).  ``A Projection Approach For Multiple Monotone Regression.  ArXiv:1911.07553.

\item {Linero, A.R.} and {Yang, Y.} (2018).  ``Bayesian Regression Tree Ensembles that Adapt to Smoothness and Sparsity.''   {\sl J. R. Statist. Soc. B} \,{\bf 80} 1087--1110.

\item {Mammen, E.} (1991).  ``Estimating a smooth monotone regression function.''  {\sl Ann. Statist.} {\bf 19} 724-740.

\item {Meyer, M. C.}, {Hackstadt, A. J.} and {Hoeting, J. A.} (2011). Bayesian estimation and inference for generalised partial linear models using shape-restricted splines.''  {\sl J. Nonparam. Statist.} {\bf 23} 867--884.

\item {Neelon, B.} and {Dunson, D.B.} (2004).  ``Bayesian isotonic regression and trend analysis.''  {\sl Biometrics} {\bf 60} 177--191. 
 
\item {Ramsay, J.O.} (1998).  ``Estimating smooth monotone functions.''  {\sl J. R. Statist. Soc. B} \, {\bf 60} 365--375.

\item {Rockova, V.} and {Saha, E.}  (2019).  ``On Theory for BART.''  {\sl Proceedings of the $22^{nd}$ International Conference on Artificial Intelligence and Statistics} {\bf 89}  2839--2848.

\item {Rockova, V.} and {van der Pas, S.} (2020).  ``Posterior Concentration for Bayesian Regression Trees and Forests.''  {\sl Ann. Statist.}  {\bf 48} 2108--2131.

\item {Saarela, O.} and {Arjas, E.} (2011). A method for Bayesian monotonic multiple regression.''  {\sl Scand. J. Statist.} {\bf 38} 499--513.

\item {Salomond, J.} (2014).   ``Concentration rate and consistency of the
posterior distribution for selected priors under monotonicity constraints.''   {\sl Electron. J. Statist.} 
{\bf 8}  1380--1404.

\item {Shively, T.S.}, {Sager, T.W.} and {Walker, S.G.} (2009).  ``A Bayesian approach to nonparametric monotone function estimation.''  {\sl J. R. Statist. Soc. B} \, {\bf 71} 159-175.

\item {Shively, T.S.}, {Walker, S.G.} and {Damien, P.} (2011).  ``Nonparametric function estimation subject to monotonicity, convexity and other shape constraints.''  {\sl J. Econometrics} {\bf 161} 166--81.

\item {Wang, W.} and {Welch, W.} (2018).   ``Bayesian optimization using monotonicity information and its application in machine learning hyperparameter tuning.''   {\sl Proceedings of the ICML 2018 AutoML Workshop}  1-8.

\item {Wang, X.} and {Berger, J.O.} (2016).  ``Estimating Shape Constrained Functions Using Gaussian Processes.''  {\sl  J. Uncertainty Quantification} {\bf 4} 1--25.

\item {Westling, T.} {van der Laan, M. J.} and {Carone,M.} (2020).  ``Correcting an estimator of a multivariate monotone function with isotonic regression.'' {\sl Electron. J. Statist.} {\bf14}  3032--3069.

\end{description}

\begin{acknowledgement}
The authors gratefully acknowledge support from the National Science Foundation (grants DMS-1944740 and DMS-1916233) and from a Simons Fellowship from the Isaac Newton Institute at the University of Cambridge.  We also thank the Editor, Associate Editor and referees for their many helpful suggestions.
\end{acknowledgement}

\end{document}